\documentclass[11pt,english]{article}
\usepackage[latin9]{inputenc}
\usepackage{amsmath}
\usepackage{amssymb}
\usepackage{graphicx}
\usepackage{setspace}
\usepackage{subscript}
\makeatletter

\providecommand{\tabularnewline}{\\}

\@ifundefined{date}{}{\date{}}
\usepackage{esint}
\usepackage{hyperref}
\usepackage{latexsym}
\usepackage{graphicx}\usepackage{bm}\usepackage{longtable}

\usepackage{xcolor}

\@addtoreset{equation}{section}

\makeatother

\usepackage{babel}
\begin{document}
\title{Holographic QCD\textsubscript{3} and Chern-Simons theory from anisotropic
supergravity}
\maketitle
\begin{center}
Si-wen Li\footnote{Email: siwenli@dlmu.edu.cn}, Sen-kai Luo\footnote{Email: luosenkai@dlmu.edu.cn},
Ya-qian Hu\footnote{Email: xmlhqq@dlmu.edu.cn},
\par\end{center}

\begin{center}
\emph{Department of Physics, School of Science,}\\
\emph{Dalian Maritime University, }\\
\emph{Dalian 116026, China}\\
\par\end{center}

\vspace{8mm}

\begin{abstract}
Based on the gauge-gravity duality, we study the three-dimensional
QCD (QCD\textsubscript{3}) and Chern-Simons theory by constructing
the anisotropic black D3-brane solution in IIB supergravity. The deformed
bulk geometry is obtained by performing a double Wick rotation and
dimension reduction which becomes an anisotropic bubble configuration
exhibiting confinement in the dual theory. And its anisotropy also
reduces to a Chern-Simons term due to the presence of the dissolved
D7-branes or the axion field in bulk. Using the bubble geometry, we
investigate the the ground-state energy density, quark potential,
entanglement entropy and the baryon vertex according to the standard
methods in the AdS/CFT dictionary. Our calculation shows that the
ground-state energy illustrates degenerate to the Chern-Simons coupling
coefficient which is in agreement with the properties of the gauge
Chern-Simons theory. The behavior of the quark tension, entanglement
entropy and the embedding of the baryon vertex further implies strong
anisotropy may destroy the confinement. Afterwards, we additionally
introduce various D7-branes as flavor and Chern-Simons branes to include
the fundamental matter and effective Chern-Simons level in the dual
theory. By counting their orientation, we finally obtain the associated
topological phase in the dual theory and the critical mass for the
phase transition. Interestingly the formula of the critical mass reveals
the flavor symmetry, which may relate to the chiral symmetry, would
be restored if the anisotropy increases greatly. As all of the analysis
is consistent with characteristics of quark-gluon plasma, we therefore
believe our framework provides a remarkable way to understand the
features of Chern-Simons theory, the strong coupled nuclear matter
and its deconfinement condition with anisotropy.
\end{abstract}
\newpage{}

\tableofcontents{}

\section{Introduction}

While quantum chromodynamics (QCD) is the underlying theory to describe
the strong interaction, it is usually very difficult to solve at low
energy due to its asymptotic freedom, especially in the dense matter
with finite temperature. Therefore it provides motivation to study
the dynamics of strongly coupled non-Abelian gauge field theory via
the gauge-gravity duality as an alternative option \cite{key-1,key-2}.
On the other hand, the heavy-ion collision (HIC) experiments show
that quark-gluon plasma (QGP) created in the collision is strongly
coupled \cite{key-3,key-4} and anisotropic \cite{key-5,key-6,key-7,key-8},
hence constructing the type IIB supergravity in order to investigate
the anisotropic and strongly coupled QGP or Yang-Mills theory through
gauge-gravity duality is naturally significant \cite{key-a1,key-a2,key-a3}
since, as it has been well-known, the most famous example in the gauge-gravity
duality is the corresponding between four-dimensional $\mathcal{N}=4$
$SU\left(N_{c}\right)$ super Yang-Mills theory on $N_{c}$ D3-branes
and type IIB super string theory on $\mathrm{AdS_{5}}\times S^{5}$.

A remarkable work in the top-down holographic approach to study the
anisotropy in gauge theory is \cite{key-9} in which the black D3-brane
solution in type IIB supergravity is anisotropic due to the presence
of the axion field or dissolved D7-branes in the bulk. Following the
AdS/CFT dictionary, the thermodynamics and transport properties in
an anisotropic plasma are explored holographically by the gravity
solution in \cite{key-9} which attracts many interests \cite{key-10,key-11}.
In particular, another concern in \cite{key-9} is that the presented
axion field leads to a theta term $\theta\int F\wedge F$ in the dual
theory and the $\theta$ parameter is spatially dependent. Since $\theta$-dependence
involves the topological property in gauge theory, it also gets many
attentions in theoretical and phenomenological researches \cite{key-12}.
Although the experimental value of the $\theta$ parameter is very
small, it may influence many observable effects in gauge theories
e.g. the deconfinement phase transition \cite{key-13,key-14}, the
glueball spectrum \cite{key-16}, the CP violation in hot QCD \cite{key-17,key-18},
the chiral magnet effect \cite{key-19,key-20}, the large N limit
\cite{key-15} and its holographic correspondence \cite{key-21,key-22,key-23,key-24}.
Accordingly, the holographic duality proposed in \cite{key-9} becomes
a topical issue at one stage.

Keeping these in hand, in this work, we would like to study the holographic
duality between the three-dimensional QCD (QCD\textsubscript{3})
and Chern-Simons theory based on \cite{key-9}. As the $\theta$ parameter
in the framework of \cite{key-9} linearly depends on one of the three
spatial coordinates, integrating by parts, one can get a three-dimensional
Chern-Simons term as $\theta\int F\wedge F\sim\int dz\wedge\mathrm{Tr}\left(A\wedge F+\frac{2}{3}A^{3}\right)$
which accordingly is the part of the motivation for this work. Furthermore,
QCD\textsubscript{3} or the Chern-Simons theory involving fundamental
matters with $N_{f}$ flavors and their large N \textquoteright t
Hooft limit are also interesting topics especially in three-dimensional
case \cite{key-25,key-26,key-27,key-28,key-29,key-30,key-31}, thus
including flavors would also be our concern in this project. And the
presented anisotropy might be more closed to the realistic physical
situation in some materials.

However one of the key points here is to find a scheme to combine
the gravity system in \cite{key-9} with the three-dimensional theory
in holography. Fortunately the answer could be found in the famous
\cite{key-32,key-33} which provides the compactification in the D3-branes
system in order to obtain a three-dimensional non-supersymmetric and
non-conformal gauge theory, as it is successfully performed in the
D4/D8 approach \cite{key-34}. So by imposing the compactification
method in \cite{key-32,key-33} to the supergravity system in \cite{key-9},
in this work we first obtain the bubble configuration of the bulk
geometry which could be remarkably analytical if we take the compactification
limit (i.e. the size of the compactification direction vanishes).
Since the bubble configuration does not have a horizon, the dual theory
is at zero temperature limit. And we examine the dual theory by introducing
a probe D3-brane at the holographic boundary which exactly exhibits
a Yang-Mills plus Chern-Simons theory as it is expected. A notable
feature in our holographic setup is that the Chern-Simons level is
naturally identified to the number of the D7-branes dissolved in the
bulk which is automatically quantized. And we believe this provides
a holographic proof to the quantization of the Chern-Simons level. 

Afterwards, some of the observables are investigated by using the
standard method according to the AdS/CFT dictionary in the bubble
configuration of the bulk, specifically they are the ground-state
energy density, quark potential, entanglement entropy and the baryon
vertex. To simplify the calculation, we consider that the size of
the compacted direction trends to be vanished in the bulk geometry
throughout this work, so that the dual theory would become exactly
three-dimensional. Then our results show that the ground-state energy
is degenerate to the Chern-Simons coupling coefficient which is in
agreement with the properties of the gauge transformation in the gauge
Chern-Simons theory \cite{key-35}. Besides, the behaviors of the
quark potential and entanglement entropy depending on the position
of the fundamental string or ``slab'' resultantly reveal that the
confinement may be destroyed in hadron if the anisotropy becomes strong
enough, because the entanglement entropy may also be a characteristic
tool to detect the confinement \cite{key-36,key-37,key-38,key-39}.
Moreover, we introduce a wrapped D5-brane on $S^{5}$ as the baryon
vertex \cite{key-40} in this geometry and study its embedding configuration
as \cite{key-44}. The numerical calculation confirms the wrapped
configuration of the baryon vertex in this system and the D-brane
force illustrates the bottom of the bulk is the stable position of
a baryon vertex as it is expected to minimize its energy. Interestingly,
the numerical calculation also displays the wrapped baryon vertex
trends to become unwrapped by the increasing of the anisotropy which
means the baryon vertex may not stably exist if the anisotropy becomes
very large. And it is seemingly consistent with the analysis of the
quark potential and entanglement entropy with respect to the confinement
in this system i.e. strong anisotropy may destroy the confinement.

Last but not least, to explore the Chern-Simons topological feature
involving the flavors in the dual theory, following \cite{key-45,key-46,key-47},
various D7-branes as flavor and Chern-Simons branes are introduced
into the bulk bubble configuration as probes. In a transverse plane,
the vacuum configuration of the D7-branes, which means the embedding
function minimizes the energy of the D7-branes, is numerically evaluated
and the calculation shows the vacuum structure is shifted by the presence
of the axion field in the bulk. To further consider the spontaneous
breaking of the flavor symmetry $U\left(N_{f}\right)$, we separate
coincident $p$ of $N_{f}$ flavor branes living into the upper part
of the transverse plane and the other coincident $N_{f}-p$ flavor
branes living into the lower part of the plane while they extend to
a same position at the holographic boundary. Therefore the interpretation
of such configuration could be that the flavor symmetry $U\left(N_{f}\right)$
spontaneously breaks down into $U\left(p\right)\times U\left(N_{f}-p\right)$
at low energy in dual theory. Taking into account the contribution
of the orientation of the flavor and Chern-Simons branes, we can get
an effective flavor-dependent Chern-Simons level. Then evaluating
the total energy including both flavor and Chern-Simons branes by
counting the orientation in the effective Chern-Simons level, the
associated topological phases in the dual theory can be obtained.
A noteworthy conclusion here is that the total energy including flavor
and Chern-Simons branes illustrates the topological phase transition
may occur at a critical flavor mass $m^{*}$ which decreases due to
the presence of the anisotropy or the axion field in the bulk geometry.
By analyzing the phase diagram, it seemingly means the broken flavor
symmetry $U\left(p\right)\times U\left(N_{f}-p\right)$, which may
relates to the chiral symmetry, would become restored to $U\left(N_{f}\right)$
if the anisotropy becomes sufficiently strong. And this behavior is
also predicted by the numerical evaluation of the embedding of the
flavor branes since the two branches of the flavor branes trend to
become coincident when the anisotropy becomes large. Altogether, this
framework may provide a holographic way to study the behavior of metastable
vacua in large N QCD\textsubscript{3} with a Chern-Simons term \cite{key-48}
and its deconfined condition with anisotropy, although the anisotropy
is expected to be small for the numerical calculations in this project.

The outline of this manuscript is as follows. In Section 2, we briefly
review the anisotropic black brane solution in the type IIB supergravity,
then give our holographic setup to this work. In Section 3, we calculate
several observables with respect to the constructed bulk geometry.
In Section 4, we discuss the embedding of the flavor and Chern-Simons
branes. In Section 5, we analyze the corresponding topological phase
and its associated phase transition in holography. Summary and discussion
are given in the final section. In addition, we list the relevant
parts of the functions presented in the bulk geometry in the appendix
which would be very useful to this work.

\section{Holographic setup}

\subsection{Review of the anisotropic solution in type IIB supergravity}

In this subsection, we review and collect the relevant content of
the anisotropic solution in ten-dimensional type IIB supergravity
in \cite{key-9}. The remarkable anisotropic solution describes the
bulk dynamics of $N_{c}$ D3-branes with $N_{\mathrm{D7}}$ D7-branes
dissolved in the spacetime in the large $N_{c}$ limit and the D-brane
configuration is given in Table \ref{tab:1}. 
\begin{table}[h]
\begin{centering}
\begin{tabular}{|c|c|c|c|c|c|c|}
\hline 
Black brane background & $t$ & $x$ & $y$ & $z$ & $u$ & $\Omega_{5}$\tabularnewline
\hline 
\hline 
$N_{c}$ D3-branes & - & - & - & - &  & \tabularnewline
\hline 
$N_{\mathrm{D7}}$ D7-branes & - & - & - &  &  & -\tabularnewline
\hline 
\end{tabular}
\par\end{centering}
\caption{\label{tab:1} The configuration of the D-branes in the black brane
background. ``-'' represents the D-brane extends along the direction.}
\end{table}
 As our concern would be the holographic duality, let us start with
the type IIB supergravity action in string frame,

\begin{equation}
S_{\mathrm{IIB}}=\frac{1}{2\kappa_{10}^{2}}\int d^{10}x\sqrt{-g}\left[e^{-2\phi}\left(\mathcal{R}+4\partial_{M}\phi\partial^{M}\phi\right)-\frac{1}{2}F_{1}^{2}-\frac{1}{4\cdot5!}F_{5}^{2}\right],\label{eq:1}
\end{equation}
where the index $M$ runs over 0 to 9, $\kappa_{10}$ is the ten-dimensional
gravitational coupling constant $2\kappa_{10}^{2}=\left(2\pi\right)^{7}l_{s}^{8}$.
To obtain an anisotropic solution, the associated equation of motion
to (\ref{eq:1}) can be solved by the following anisotropic ansatz
in string frame,

\begin{align}
ds^{2} & =\frac{L^{2}}{u^{2}}\left(-\mathcal{F}\mathcal{B}dt^{2}+dx^{2}+dy^{2}+\mathcal{H}dz^{2}+\frac{du^{2}}{\mathcal{F}}\right)+L^{2}\mathcal{Z}d\Omega_{5}^{2},\nonumber \\
F_{1} & =d\chi,\ \chi=az,\ F_{5}=dC_{4}=\frac{4}{L}\left(\Omega_{S^{5}}+\star\Omega_{S^{5}}\right),\nonumber \\
\mathcal{H} & =e^{-\phi},\ \mathcal{Z}=e^{\frac{1}{2}\phi},\label{eq:2}
\end{align}
where $\chi,\phi,\Omega_{S^{5}}$ refers to the axion, dilaton and
the unit volume form of a five-sphere $S^{5}$. The parameters in
the solution are given as follows, 

\begin{equation}
L^{4}=4\pi g_{s}N_{c}l_{s}^{4}=\lambda l_{s}^{4},a=\frac{\lambda n_{\mathrm{D7}}}{4\pi N_{c}},\label{eq:3}
\end{equation}
where $L,g_{s},\lambda$ represents the radius of the bulk, the string
coupling and the 't Hooft coupling constant respectively. The solution
(\ref{eq:2}) describes the black branes with a horizon at $u=u_{H}$
and the anisotropy in $z$ direction. There would not be new field
in the boundary which is located at $u=0$ because the $N_{\mathrm{D7}}$
D7-branes do not extend along the holographic direction $u$. Since
the dynamic of the axion $\chi$, which magnetically couples to $N_{\mathrm{D7}}$
D7-branes, is taken into account, it is clear that the supergravity
solution (\ref{eq:2}) includes the backreaction of $N_{\mathrm{D7}}$
D7-branes to the D3-brane bulk geometry. We note that the $N_{\mathrm{D7}}$
D7-branes are distributed along $z$ direction with the constant distribution
density $n_{\mathrm{D7}}=dN_{\mathrm{D7}}/dz$ according to the solution
for $\chi$. So once the backreaction of $N_{\mathrm{D7}}$ D7-branes
to the background geometry is included, it implies that $N_{\mathrm{D7}}/N_{c}$
is fixed in the large $N_{c}$ limit. 

The regular functions $\mathcal{F},\mathcal{B},\phi$ in (\ref{eq:2})
depend on the holographic coordinate $u$ which must be determined
by their equations of motion. However they are non-analytical in general.
In order to avoid the conical singularities in the bulk, the Euclidean
version of the bulk metric near the horizon,

\begin{equation}
ds_{E}^{2}\simeq\frac{1}{u_{H}^{2}}\left[\mathcal{F}_{1}\left(u_{H}\right)\mathcal{B}\left(u_{H}\right)\left(u-u_{H}\right)\left(dt_{E}\right)^{2}+\frac{du^{2}}{\mathcal{F}_{1}\left(u-u_{H}\right)}\right],\ \mathcal{F}_{1}=-\frac{d\mathcal{F}}{du},
\end{equation}
must impose the period $\delta t_{E}$ to be $2\pi$. Hence it reduces
to the formula of the Hawking temperature $T$ as,

\begin{equation}
\delta t_{E}=\frac{4\pi}{\mathcal{F}_{1}\left(u_{H}\right)\sqrt{\mathcal{B}_{H}}}=\frac{1}{T}.
\end{equation}
Suppose the temperature is sufficiently large $T\rightarrow\infty$
(or equivalently $\delta t_{E}\rightarrow0$), the functions $\mathcal{F},\mathcal{B},\phi$
can be analytically written as the series of $a$ which are given
in the Appendix. We note this high-temperature analysis would be remarkably
useful in the following sections of this work.

\subsection{Construction for the 2+1 dimensional theory}

\begin{figure}
\begin{centering}
\includegraphics[scale=0.25]{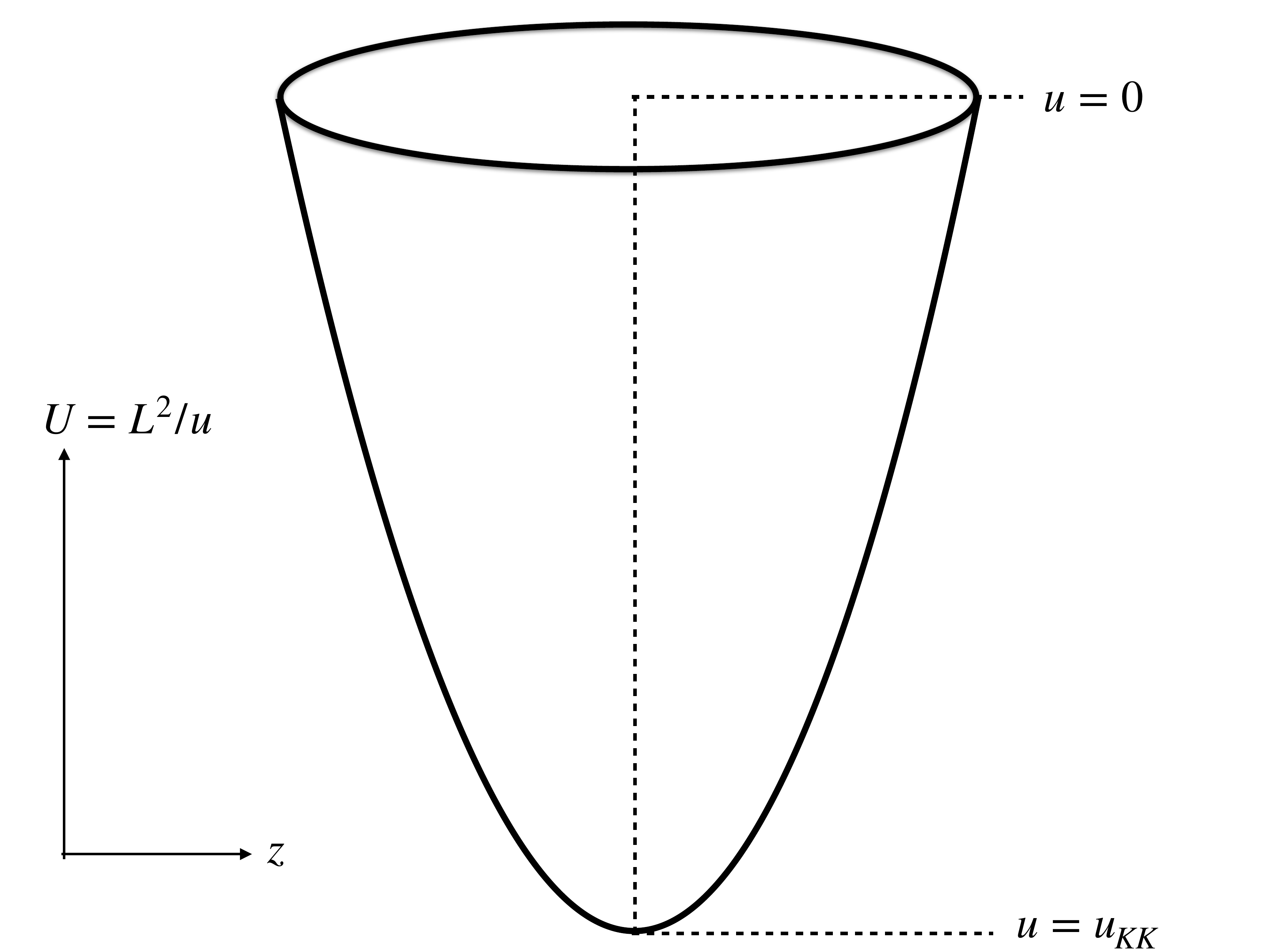}
\par\end{centering}
\caption{\label{fig:1} The bubble geometry of the $N_{c}$ D3-brane. The bulk
ends at $u=u_{KK}$ and the holographic boundary is located at $u=0$.
We note that the standard holographic radius coordinate is in fact
defined as $U=L^{2}/u$, so the bubble configuration is exactly illustrated
in the $\left\{ U,z\right\} $ plane.}

\end{figure}

As it is known that the type IIB supergravity theory holographically
corresponds to the $\mathcal{N}=4$ super Yang-Mills theory on D3-brane,
it would be very straightforward to construct the D3-brane configuration
or the $\mathcal{N}=4$ super Yang-Mills theory in order to obtain
a non-supersymmetric and non-conformal dual theory by following the
steps in the well-known \cite{key-32,key-33}. Specifically the first
step is to take the spatial dimensions $y$ of the D3-brane to be
compactified on a circle $S^{1}$ with a period $\delta y$. Therefore
the dual theory is effectively three-dimensional below the Kaluza-Klein
energy scale defined as $M_{KK}=2\pi/\delta y$. The second step is
going to get rid of all massless fields other than the gauge fields,
which is to impose respectively the periodic and anti-periodic boundary
condition on bosonic and fermionic fields along $S^{1}$. Afterwards
the supersymmetric fermions and scalars acquire mass of order $M_{KK}$
thus they are decoupled in the low-energy dynamics. So the dual theory
below $M_{KK}$ becomes three-dimensional pure gauge theory. By keeping
these in mind, the next step is to identify the bulk geometry that
corresponds to this gauge theory. The answer can be found by interchanging
the roles of $t$ and $y$ i.e. performing a double Wick rotation
$t\rightarrow-iy,y\rightarrow-it$ to the metric presented in (\ref{eq:2}),
which is

\begin{equation}
ds^{2}=\frac{L^{2}}{u^{2}}\left(-dt^{2}+dx^{2}+\mathcal{H}dy^{2}+\mathcal{F}\mathcal{B}dz^{2}+\frac{du^{2}}{\mathcal{F}}\right)+L^{2}\mathcal{Z}d\Omega_{5}^{2},\label{eq:6}
\end{equation}
where we have renamed $y,z$ after the double Wick rotation\footnote{Performing $t\rightarrow-iy,y\rightarrow-it$ to the metric presented
in (\ref{eq:2}) reduces to 
\[
ds^{2}=\frac{L^{2}}{u^{2}}\left(-dt^{2}+dx^{2}+\mathcal{F}\mathcal{B}dy^{2}+\mathcal{H}dz^{2}+\frac{du^{2}}{\mathcal{F}}\right)+L^{2}\mathcal{Z}d\Omega_{5}^{2}.
\]
Then we rename $y,z$ by $y,z\rightarrow z,y$ in order to obtain
(\ref{eq:6}). }. We note that in this notation the axion field becomes,

\begin{equation}
\chi=ay,
\end{equation}
while $\phi,F_{5}$ remains and now $z$ is periodic as,

\begin{equation}
\delta z=\frac{4\pi}{\mathcal{F}_{1}\left(u_{KK}\right)\sqrt{\mathcal{B}\left(u_{KK}\right)}}=\frac{2\pi}{M_{KK}}.
\end{equation}
Using the formulas given in the appendix, we can obtain

\begin{equation}
M_{KK}=\frac{2}{u_{KK}}+\frac{u_{KK}}{24}\left(5\log2-2\right)a^{2}+\mathcal{O}\left(a^{4}\right).\label{eq:9}
\end{equation}
The solution (\ref{eq:6}) represents a bubble geometry of the bulk
which is anisotropic on $\left\{ x,y\right\} $ plane and defined
only for $0\leq u\leq u_{KK}$. The D-brane configuration for the
bubble solution (\ref{eq:6}) is given in Table \ref{tab:2}. 
\begin{table}
\begin{centering}
\begin{tabular}{|c|c|c|c|c|c|c|}
\hline 
Bubble background & $t$ & $x$ & $y$ & $\left(z\right)$ & $u$ & $\Omega_{5}$\tabularnewline
\hline 
\hline 
$N_{c}$ D3-branes & - & - & - & - &  & \tabularnewline
\hline 
$N_{\mathrm{D7}}$ D7-branes & - & - &  & - &  & -\tabularnewline
\hline 
\end{tabular}
\par\end{centering}
\caption{\label{tab:2} The configuration of the D-branes in the bubble solution
(\ref{eq:6}).}
\end{table}
 Here we have renamed $u_{H}$ as $u_{KK}$ since there is not a horizon
in the bulk as it is illustrated in Figure \ref{fig:1}. 
 Since the wrap factor $L^{2}/u^{2}$ never goes to zero, the dual
theory would exhibit confinement according to the behavior of the
Wilson loop in this bulk geometry. Besides, we can notice that the
gauge theory on D3-brane becomes purely three-dimensional theory if
the compactified direction $z$ shrinks to zero i.e. $\delta z\rightarrow0$
or $M_{KK}\rightarrow\infty$. And this limit exactly corresponds
to the high temperature limit in the black brane solution (\ref{eq:2})
so that in the limit of $\delta z\rightarrow0$, the functions $\mathcal{F},\mathcal{B},\phi$
in (\ref{eq:6}) are also analytical as they are given in the appendix.
Accordingly we are going to consider the case in the limit $\delta z\rightarrow0$
throughout this work since our concern is the three-dimensional dual
theory exactly\footnote{A safe statement is to further require $N_{\mathrm{D7}}/N_{c}\ll1$
in the construction of the bubble geometry since the limit of $N_{\mathrm{D7}}/N_{c}\gg1$would
lead to pure D7-brane background which may have many issues in a holographic
approach. To avoid those issues, we may consider that $N_{\mathrm{D7}}/N_{c}\ll1$
with fixed $N_{\mathrm{D7}}/N_{c}\sim a$ in the large $N_{c}$ for
the gravity background produced by multiple D-branes as in \cite{key-59,key-60}
which means the anisotropy may not be very large in this setup.}.

\subsection{The dual theory}

The dual theory with respect to the bulk solution (\ref{eq:6}) is
defined at the zero temperature limit $T\rightarrow0$ since there
is not a horizon i.e. $\delta t\rightarrow\infty$. To examine the
dual theory, let us introduce a single probe D3-brane located at the
boundary $u\rightarrow0$. As we have discussed, in the bubble geometry
(\ref{eq:6}), the low-energy modes in the dual field theory contain
the gauge field only, so the effective action for such a probe D3-brane
is given as,

\begin{equation}
S_{\mathrm{D3}}=-T_{3}\mathrm{Tr}\int d^{3}xdze^{-\phi}\sqrt{-\det\left(g_{ab}+\mathcal{F}_{ab}\right)}+\frac{1}{2}T_{3}\mathrm{Tr}\int\chi\mathcal{F}\wedge\mathcal{F},\label{eq:10}
\end{equation}
where the tension of the D3-brane is given as $T_{3}=\left(2\pi\right)^{-3}l_{s}^{-4}g_{s}^{-1}$.
And $g_{ab},\mathcal{F}_{ab}=2\pi\alpha^{\prime}F_{ab}$ is the induced
metric and the gauge field strength on the worldvolume of the D3-brane.
Assuming $\mathcal{F}_{ab}$ does not have components along $z$ and
does not depend on $z$, then the quadratic expansion of the action
(\ref{eq:10}) is

\begin{align}
S_{\mathrm{D3}} & =-\frac{1}{4}T_{3}\left(2\pi\alpha^{\prime}\right)^{2}\mathrm{Tr}\int_{\mathrm{D3}}d^{3}xdz\sqrt{-g}e^{-\phi}g^{ac}g^{bd}F_{ab}F_{cd}-\mu_{3}\left(2\pi\alpha^{\prime}\right)^{2}\mathrm{Tr}\int_{\mathrm{D3}}d\chi\wedge\omega_{3}\nonumber \\
 & =-\frac{N_{c}}{4\lambda_{3}}\mathrm{Tr}\int_{\mathbb{R}^{2+1}}d^{3}xF_{ab}^{2}-\frac{N_{\mathrm{D7}}}{4\pi}\mathrm{Tr}\int_{\mathbb{R}^{2+1}}\omega_{3},\label{eq:11}
\end{align}
where $\lambda_{3}=\lambda M_{KK}/\left(2\pi\right)$ is the three-dimensional
't Hooft coupling constant and 

\begin{equation}
\omega_{3}=A\wedge dA+\frac{2}{3}A\wedge A\wedge A,
\end{equation}
is the Chern-Simons three-form. We use $A$ to denote the gauge potential
and have imposed the boundary value $\phi_{\mathrm{bdry}}=0,\mathcal{F}_{\mathrm{bdry}}=\mathcal{B}_{\mathrm{bdry}}=1$
to (\ref{eq:11}). Clearly the dual theory on the D3-brane is effectively
three-dimensional Yang-Mills plus Chern-Simons theory below the energy
scale $M_{KK}$. Notice that the dual theory is expected to be purely
three-dimensional theory if we take the limit $\delta z\rightarrow0$
or equivalently $M_{KK}\rightarrow\infty$.

It is remarkable to notice that in this holographic setup, the level
number of the Chern-Simons term $N_{\mathrm{D7}}$ is integer automatically
since the level number is exactly the number of D7-branes in the gravity
side. This leads to a proof of the quantization of the Chern-Simons
level via holography. On the other hand, when the backreaction of
the $N_{\mathrm{D7}}$ D7-branes is included in the bulk, the dual
theory is equivalently a topological massive theory. This can be confirmed
once we derive the formula of the propagator with respect to action
(\ref{eq:11}) which is \cite{key-35},

\begin{equation}
\Delta_{ab}=\frac{p^{2}\eta_{ab}-p_{a}p_{b}+i\kappa g_{YM}^{2}\epsilon_{abc}p^{c}}{p^{2}\left(p^{2}-\kappa^{2}g_{YM}^{2}\right)}+\mathrm{gauge\ fixing\ terms},
\end{equation}
where $\kappa=N_{\mathrm{D7}}/\left(4\pi\right)$ and we use $p$
to denote the momentum in 2+1 dimensional spacetime. Therefore, the
propagator defines the topological mass of the gauge field via the
pole $p^{2}=\kappa^{2}g_{YM}^{2}$ which is determined by the numbers
of D7-branes. Thus the presented D7-branes involve the topological
properties of the dual theory and, we will see, they contribute to
the various vacuum configurations in the dual theory.

\section{Observables}

In this section, we will extract relevant information on the physics
of the Yang-Mills theory with a Chern-Simons term dual to the anisotropic
background given in (\ref{eq:6}). Using the standard holographic
methods, we will focus on the ground-state energy, quark potential,
entanglement entropy and baryon vertex in this system. 

\subsection{The ground-state energy}

\begin{figure}
\begin{centering}
\includegraphics[scale=0.6]{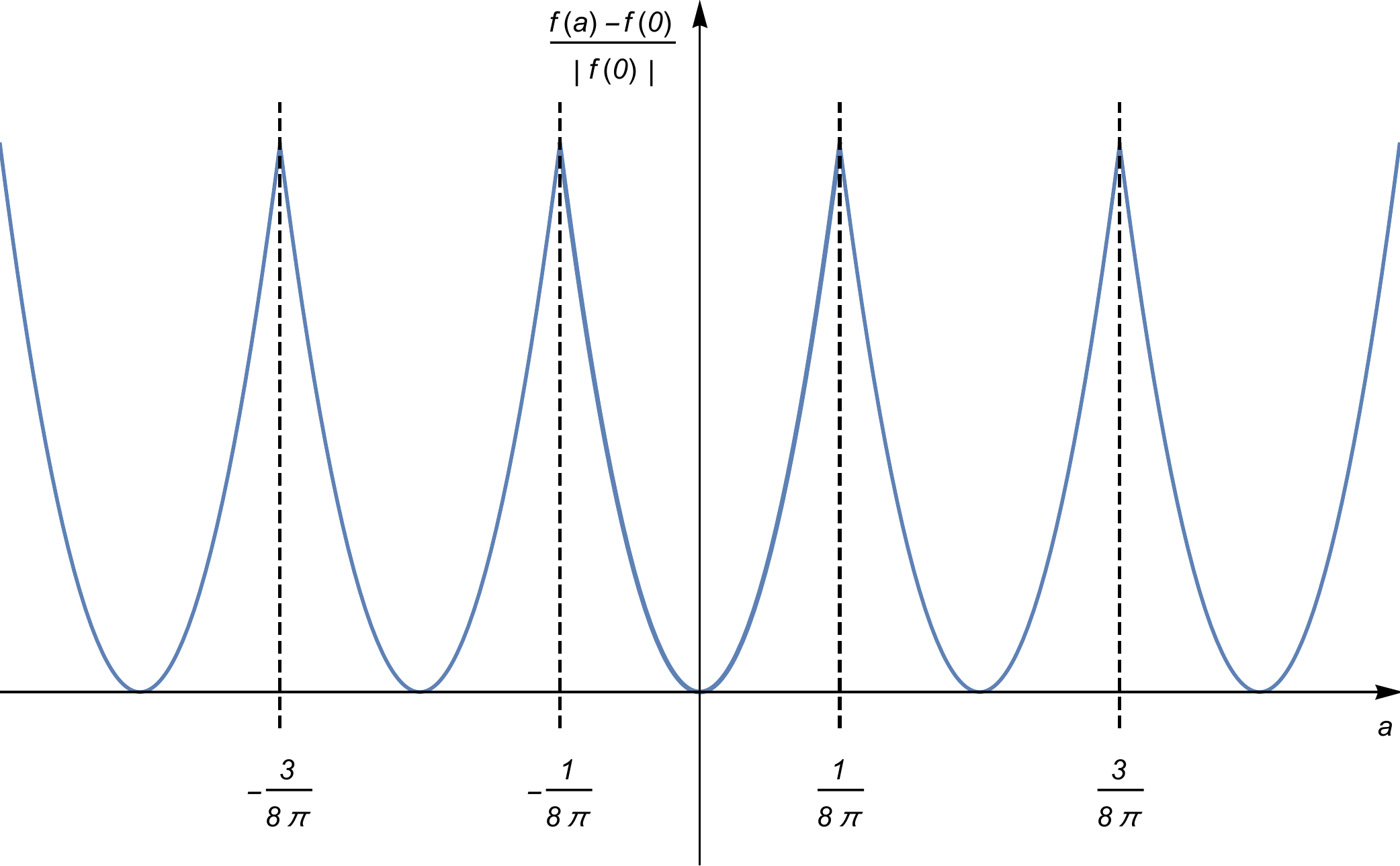}
\par\end{centering}
\caption{\label{fig:2} The vacuum energy for distinct $\kappa$. For example
for $\kappa=0$, we have $a\in\left[-\frac{1}{8\pi},\frac{1}{8\pi}\right]$;
for $\kappa=\frac{1}{4\pi}$, we have $a\in\left[\frac{1}{8\pi},\frac{3}{8\pi}\right]$;
for $\kappa=-\frac{1}{4\pi}$, we have $a\in\left[-\frac{3}{8\pi},-\frac{1}{8\pi}\right]$
and so on.}

\end{figure}

In the holographic dictionary, one of the basic entries is the relation
between the renormalized on-shell supergravity action and partition
function of the dual field theory \cite{key-1,key-2,key-32,key-33}.
Therefore the ground-state energy density $f$ of the dual field theory
can be obtained through the relation

\begin{equation}
Z=e^{-V_{3}f}=e^{-S_{E,\mathrm{on-shell}}^{ren}},\label{eq:14}
\end{equation}
where $V_{3}$ refers to the infinite three-dimensional Euclidean
spacetime volume. $S_{E,\mathrm{on-shell}}^{ren}$ is the renormalized
on-shell action for type IIB supergravity, which is given by

\begin{equation}
S_{E,\mathrm{on-shell}}^{ren}=S_{\mathrm{IIB}}^{E}+S_{\mathrm{GH}}+S_{\mathrm{CT}},\label{eq:15}
\end{equation}
where $S_{\mathrm{GH}},S_{\mathrm{CT}}$ refers to the Gibbons-Hawking
term and holographic counterterm for the type IIB supergravity. $S_{\mathrm{IIB}}^{E}$
refers to the Euclidean version of (\ref{eq:1}) which in Einstein
frame is given as,

\begin{equation}
S_{\mathrm{IIB}}^{E}=-\frac{1}{2\kappa^{2}}\int_{\mathcal{M}}d^{10}x\sqrt{g}\left[\mathcal{R}-\frac{1}{2}\partial_{M}\phi\partial^{M}\phi-\frac{1}{2}e^{2\phi}F_{1}^{2}-\frac{1}{4\cdot5!}F_{5}^{2}\right].\label{eq:16}
\end{equation}
Since only $F_{5}$ has components on $\Omega_{5}$, the onshell action
$S_{\mathrm{IIB}}^{E}$ can be integrated out over $\Omega_{5}$ to
become an effective five-dimensional action as,

\begin{equation}
S_{\mathrm{IIB}}^{E}=\frac{1}{2\kappa_{5}^{2}}\int_{\mathcal{M}}d^{5}x\sqrt{g}\left(\mathcal{R}^{\left(5\right)}-2\Lambda-\frac{1}{2}\partial_{M}\phi\partial^{M}\phi-\frac{1}{2}e^{2\phi}\partial_{M}\chi\partial^{M}\chi\right),\label{eq:17}
\end{equation}
where $M$ runs over 0 to 4. The cosmological constant is $\Lambda=-6/L^{2}$,
$\mathcal{R}^{\left(5\right)}$ is the five-dimensional scalar curvature
and $\kappa_{5}$ is the five-dimensional gravitational coupling constant.
The action (\ref{eq:17}) is nothing but the five-dimensional axion-dilaton-gravity
action. Hence the holographic counterterm can be chosen as \cite{key-9},

\begin{equation}
S_{\mathrm{CT}}=-\frac{1}{\kappa_{5}^{2}}\int_{\partial\mathcal{M}}d^{4}x\sqrt{h}\left(3-\frac{1}{8}e^{2\phi}h^{\mu\nu}\partial_{\mu}\chi\partial_{\nu}\chi\right)+\log v\int_{\partial\mathcal{M}}d^{4}x\sqrt{h}\mathcal{A}-\frac{1}{4}\left(c_{\mathrm{sch}}-1\right)\int_{\partial\mathcal{M}}d^{4}x\mathcal{A},\label{eq:18}
\end{equation}
where $h_{\mu\nu}$ refers to the boundary metric and $\mathcal{A}\left(h_{\mu\nu},\phi,\chi\right)$
refers to the conformal anomaly in the axion-dilaton-gravity system
\footnote{The counterterm in axion-dilaton-gravity system can also be found
in \cite{key-49,key-50} }. In this sense, the metric near the boundary is required to take
the form as,
\begin{equation}
ds^{2}=\frac{dv^{2}}{v^{2}}+h_{\mu\nu}dx^{\mu}dx^{\nu}.
\end{equation}
in which the coordinate $v$ is the standard Fefferman-Graham (FG)
coordinate. The relation between the coordinate $u$ presented in
(\ref{eq:6}) and the FG coordinate is collected as,

\begin{equation}
u=v+\frac{a^{2}}{12}v^{3}+\mathcal{O}\left(v^{5}\right),v=u-\frac{a^{2}}{12}u^{3}+\mathcal{O}\left(u^{5}\right).
\end{equation}
We also note that the standard Gibbons-Hawking term is given by the
trace of the extrinsic curvature $K$ of the boundary as

\begin{equation}
S_{\mathrm{GH}}=\frac{1}{\kappa_{5}^{2}}\int_{\partial\mathcal{M}}\sqrt{h}K.\label{eq:21}
\end{equation}
Plugging (\ref{eq:17}) (\ref{eq:18}) (\ref{eq:21}) into (\ref{eq:15})
and using the relation (\ref{eq:14}), the resultant free energy density
of the dual theory is computed up to $\mathcal{O}\left(a^{2}\right)$
as,

\begin{equation}
f=-\frac{M_{KK}^{3}N_{c}^{2}}{64\pi}+\frac{a^{2}M_{KK}N_{c}^{2}}{64\pi}+\mathcal{O}\left(a^{4}\right).\label{eq:22}
\end{equation}
The free energy density depending on $a$ implies the ground-state
is degenerate to the Chern-Simons coupling coefficient $\kappa=N_{\mathrm{D7}}/\left(4\pi\right)$.
Since $a\sim dN_{\mathrm{D7}}/dy$ is the distribution density of
the $N_{\mathrm{D7}}$ D7-branes, the value of $a$ could be same
for different $\kappa$. Therefore the (\ref{eq:22}) describes the
vacuum energy in various branches characterized by its level number
$4\pi\kappa$. To any given interval, we fix the length $1/\left(4\pi\right)$
of $a$ for possible values of $\kappa$, then the ground-state free
energy (\ref{eq:22}), which implies the vacuum with distinct $\kappa$,
is shown in Figure \ref{fig:2}. 
 And the true vacuum should minimize the free energy (\ref{eq:22}).

Figure \ref{fig:2} also illustrates that the behavior of ground-state
energy is similar to the same observable in QCD\textsubscript{4}
with finite theta angle, especially in some holographic approaches
\cite{key-21,key-22,key-23}. The reason is as follows. First, the
black brane solution (\ref{eq:2}) and bubble solution (\ref{eq:6})
share same value of the onshell action in gravity side since the difference
here between them is just the double Wick rotation. Then, on the other
hand, the dual theory in the the black brane and bubble background
is respectively four-dimensional theta-depended gauge theory and three-dimensional
Yang-Mills-Chern-Simons given in (\ref{eq:11}). Therefore it is not
surprised that their dual theories also share similar behavior of
the ground-state energy according to the AdS/CFT dictionary (\ref{eq:14})
\footnote{One may consider the dualities in string theory to study this similarity
in another way, since the holographic investigation in \cite{key-21,key-22,key-23}
is based on the D4-brane approach in IIA string theory which is a
T-duality version of IIB string theory.}. In addition, we may find in (\ref{eq:10}) (\ref{eq:11}), the Chern-Simons
three-form comes from the integral by part to the four-dimensional
Wess-Zumino term in which the axion field plays exactly the role of
the theta term as in QCD\textsubscript{4}. Thus the similarity in
the behaviors of the ground-state energy with respect to QCD\textsubscript{3}
and QCD\textsubscript{4} may also be understood by this integral
relation.

Besides, the degeneracy to the Chern-Simons level number in the ground-state
free energy via holography may also have an interpretation in terms
of quantum field theory (QFT). As we have analyzed, the dual theory
on D3-brane is the three-dimensional Yang-Mills-Chern-Simons whose
action is given by (\ref{eq:11}). So under the local $SU\left(N_{c}\right)$
gauge transformation

\begin{equation}
A_{a}\rightarrow U^{-1}A_{a}U+U^{-1}\partial_{a}U,
\end{equation}
the Yang-Mills-Chern-Simons action $S_{\mathrm{YMCS}}$ presented
in (\ref{eq:11}) transforms as,

\begin{equation}
S_{\mathrm{YMCS}}\rightarrow S_{\mathrm{YMCS}}-8\pi^{2}\kappa W,
\end{equation}
where $W$ is the winding number given by 

\begin{equation}
W=\frac{1}{24\pi^{2}}\int d^{3}x\epsilon^{abc}\mathrm{Tr}\left(U\partial_{a}UU^{-1}\partial_{b}UU^{-1}\partial_{c}U\right).
\end{equation}
Therefore, under the gauge transformation, the partition function
given by the Euclidean path integral is

\begin{equation}
Z=\int\mathcal{D}Ae^{iS_{\mathrm{YMCS}}\left[A\right]}=\int\mathcal{D}Ae^{iS_{\mathrm{YMCS}}\left[A\right]-8i\pi^{2}\kappa W},4\pi\kappa\in\mathrm{integer},\label{eq:26}
\end{equation}
up to a phase $e^{-8i\pi^{2}\kappa W}$. Comparing (\ref{eq:26})
with the AdS/CFT dictionary (\ref{eq:14}), it means the free energy
must be degenerate to $\kappa$. This property of Chern-Simons theory
also leads to an additive renormalization condition to the renormalized
and bare Chern-Simons coupling coefficient

\begin{equation}
4\pi\kappa_{\mathrm{ren}}=4\pi\kappa_{\mathrm{bare}}+N_{c},
\end{equation}
up to the one-loop order calculation at least, according to \cite{key-35}.
It also implies the renormalized parameter (denoted by $a$ in our
system) is degenerate to the bare parameter (denoted by $\kappa$)
up to a finite integer-valued shift. 

\subsection{Wilson loop and quark potential}

In holography, the vacuum expectation value (VEV) of Wilson loop on
a contour $\mathcal{C}$ corresponds to the renormalized Nambu-Goto
on-shell action of a fundamental open string whose endpoints span
the contour $\mathcal{C}$ \cite{key-51} which is,

\begin{equation}
\left\langle W\left(\mathcal{C}\right)\right\rangle =e^{-S_{NG}}.
\end{equation}
The static quark-antiquark potential $V$ can be obtained by evaluating
the Nambu-Goto action as,

\begin{equation}
S_{NG}=-\mathcal{T}\left(V+2M_{q\bar{q}}\right),\label{eq:29}
\end{equation}
where $m$ refers to the bare mass of quark which is also the counterterm
in the Nambu-Goto action. In the background (\ref{eq:2}), the bare
mass can be evaluated by putting the fundamental open string extending
along $u$ direction. Choosing $\tau=t,\sigma=u$, the induced metric
on the worldsheet parametrized by $\left\{ \tau,\sigma\right\} $
is written as,

\begin{equation}
ds^{2}=\frac{L^{2}}{u^{2}}\left(-dt^{2}+\frac{du^{2}}{\mathcal{F}}\right).\label{eq:30}
\end{equation}
Hence the bare mass is obtained by the Nambu-Goto action with respected
to (\ref{eq:30}) as,

\begin{equation}
M_{q\bar{q}}=\frac{\sqrt{\lambda}}{4\pi}\int_{u_{0}}^{0}\frac{du}{u^{2}\sqrt{\mathcal{F}}}.\label{eq:31}
\end{equation}
Next in order to compute the quark-antiquark potential, we consider
a rectangular contour with sides of length $\mathcal{T},L$ along
$t$ and one spatial direction $x$ or $y$. Notice the background
metric is anisotropic in $\left\{ x,y\right\} $ plane, the calculation
of Wilson loop would be a little different when the string extends
along $x$ and $y$ direction.

\subsubsection*{Parallel to the D7-branes}

\begin{figure}
\begin{centering}
\includegraphics[scale=0.3]{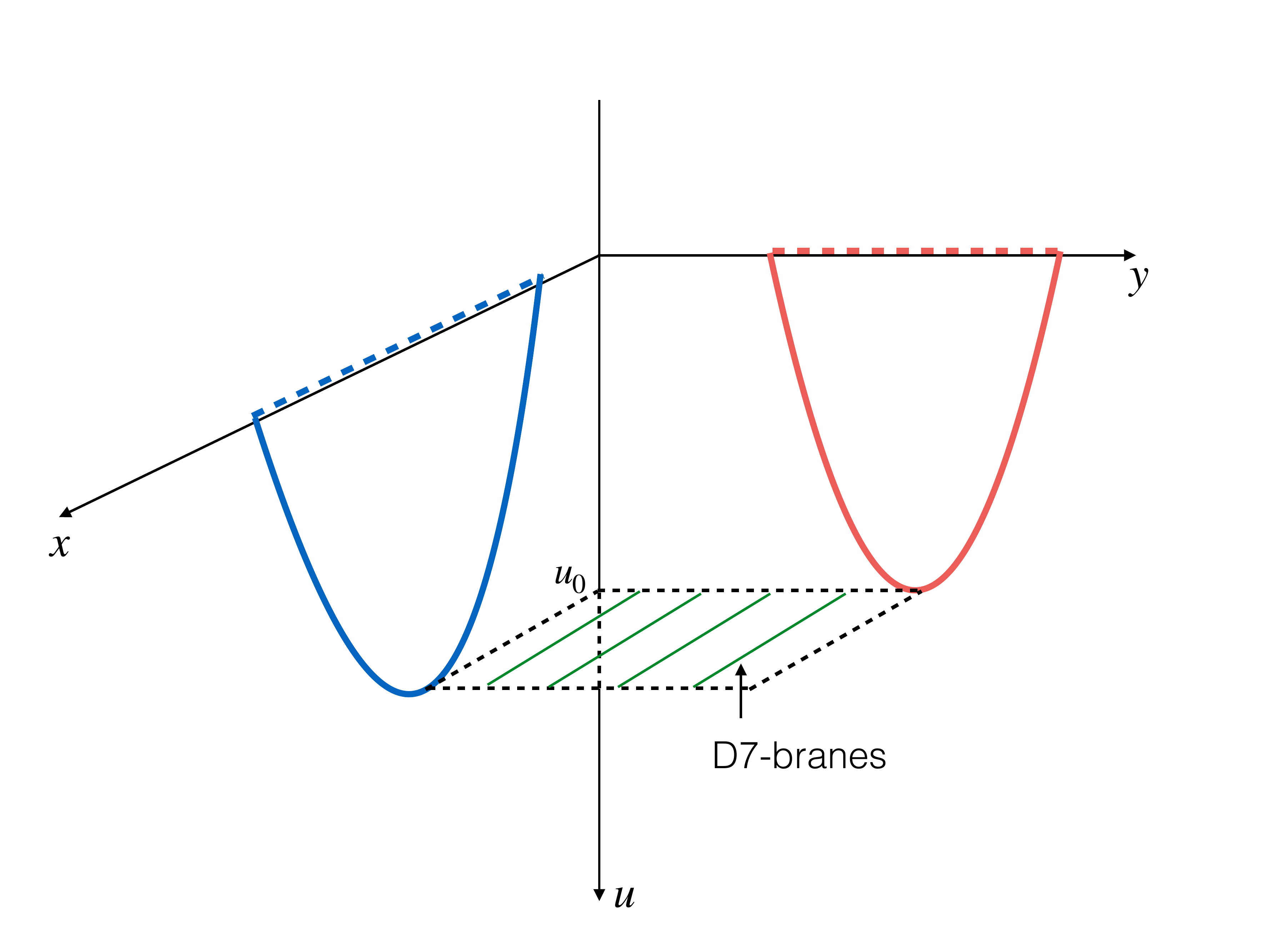}
\par\end{centering}
\caption{\label{fig:3} The configuration of the fundamental string and the
$N_{\mathrm{D7}}$ D7-branes. The blue line refers to the case that
the open string is put in $\left\{ x,u\right\} $ plane i.e. parallel
to the D7-branes. The red line refers to the case that the open string
is put in $\left\{ y,u\right\} $ plane i.e. vertical to the D7-branes.
The D7-branes are represented by the green lines.}

\end{figure}

\begin{figure}
\begin{centering}
\includegraphics[scale=0.43]{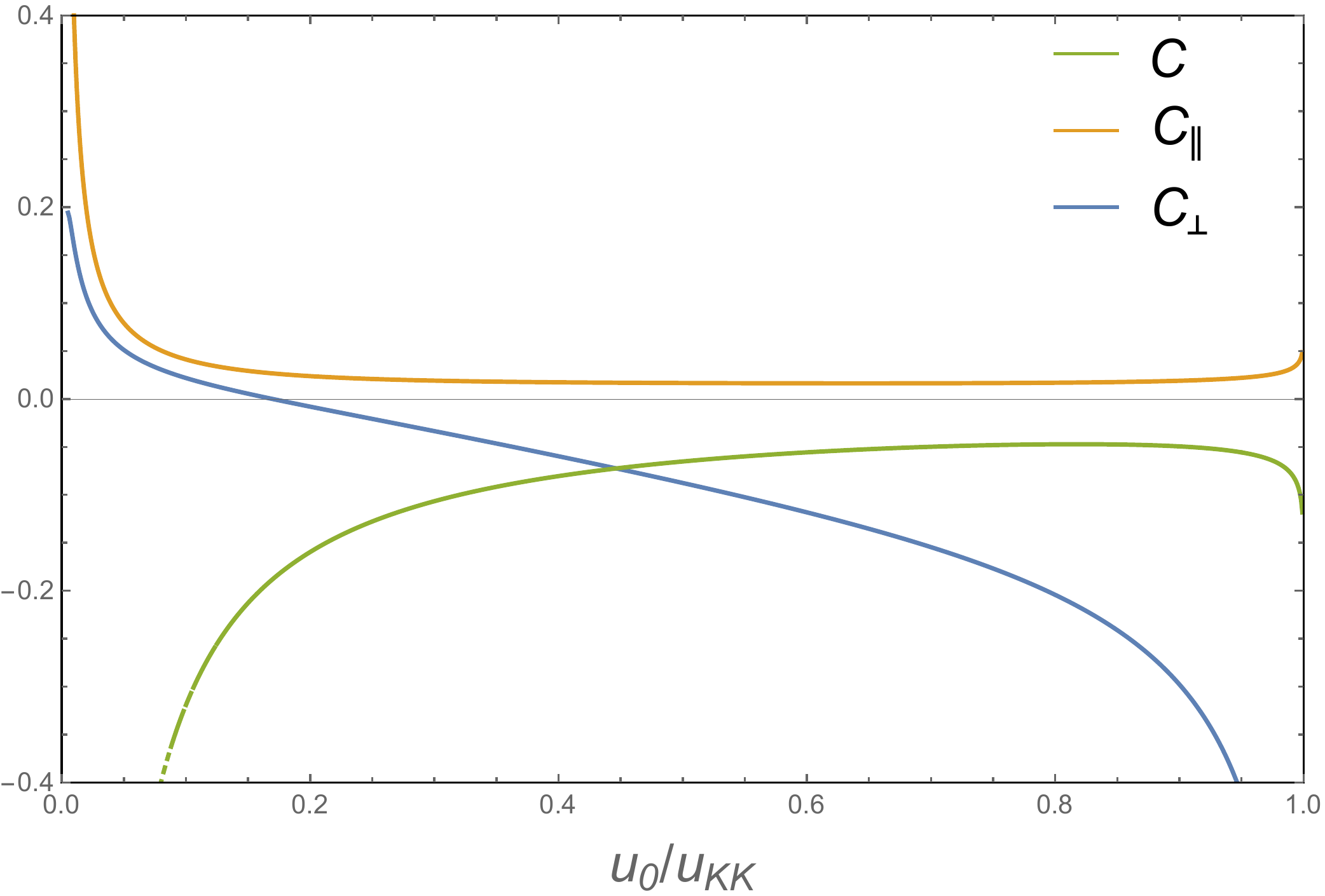}
\par\end{centering}
\caption{\label{fig:4} The numerical constants $\mathcal{C},\mathcal{C}_{\Vert},\mathcal{C}_{\bot}$
as functions of $u_{0}$.}

\end{figure}

Let us first consider the situation that the open string extends parallel
to the D7-branes as it is illustrated in Figure \ref{fig:3}. 
 Taking the static gauge as $\tau=t\in\left[0,\mathcal{T}\right],\sigma=x\in\left[-\frac{R}{2},\frac{R}{2}\right],u=u\left(x\right)$,
the Nambu-Goto action is given as,

\begin{align}
S_{NG}^{\Vert} & =-\frac{1}{2\pi\alpha^{\prime}}\int d\tau d\sigma\sqrt{-g_{\tau\tau}g_{\sigma\sigma}}\nonumber \\
 & =-\frac{\mathcal{T}}{2\pi\alpha^{\prime}}\int dx\sqrt{-g_{00}\left[g_{xx}+g_{uu}u^{\prime}\left(x\right)^{2}\right]}\nonumber \\
 & =-\frac{L^{2}\mathcal{T}}{2\pi\alpha^{\prime}}\int\frac{1}{u^{2}}\sqrt{1+\frac{u^{\prime2}}{\mathcal{F}}}dx,\label{eq:32}
\end{align}
where the derivatives `` $^{\prime}$ '' are with respect to $x$.
Notice that the associated Hamiltonian to (\ref{eq:32}) is conserved
i.e. a constant since the Lagrangian presented in (\ref{eq:32}) does
not depend on $x$ explicitly. Accordingly we can reach

\begin{align}
\mathcal{H} & =u^{\prime}\frac{\partial\mathcal{L}}{\partial u^{\prime}}-\mathcal{L}\nonumber \\
 & =\frac{L^{2}\mathcal{T}}{2\pi\alpha^{\prime}}\frac{1}{u^{2}\sqrt{1+\frac{u^{\prime2}}{\mathcal{F}}}}=\frac{L^{2}\mathcal{T}}{2\pi\alpha^{\prime}}\times\mathrm{const}.
\end{align}
Imposing the condition $u^{\prime}\left(x\right)\big|_{u=u_{0}}=0$,
the Hamiltonian reduces to

\begin{equation}
\frac{1}{u^{2}\sqrt{1+\frac{u^{\prime2}}{\mathcal{F}}}}=\frac{1}{u_{0}^{2}},
\end{equation}
which is equivalent to

\begin{equation}
\frac{du}{dx}=\sqrt{\mathcal{F}\left(\frac{u_{0}^{4}}{u^{4}}-1\right)}.\label{eq:35}
\end{equation}
Plugging (\ref{eq:35}) into (\ref{eq:32}), the Nambu-Goto action
is written as,

\begin{equation}
S_{NG}^{\Vert}=-\frac{L^{2}\mathcal{T}}{\pi\alpha^{\prime}}\int_{u_{0}}^{0}\frac{u_{0}^{2}}{u^{2}}\frac{du}{\sqrt{\mathcal{F}\left(u_{0}^{4}-u^{4}\right)}}.\label{eq:36}
\end{equation}
So according to (\ref{eq:29}), the quark-antiquark potential can
be obtained by subtracting (\ref{eq:31}) from (\ref{eq:36}) as,

\begin{align}
V^{\Vert} & =\frac{L^{2}\mathcal{T}}{\pi\alpha^{\prime}}\int_{u_{0}}^{0}\frac{1}{u^{2}\sqrt{\mathcal{F}}}\left[\frac{u_{0}^{2}}{\sqrt{\left(u_{0}^{4}-u^{4}\right)}}-1\right]\nonumber \\
 & =\lambda^{1/2}M_{KK}\mathcal{C}\left(u_{0}\right)+\frac{a^{2}}{M_{KK}}\mathcal{C}^{\Vert}\left(u_{0}\right)+\mathcal{O}\left(a^{4}\right),\label{eq:37}
\end{align}
where we have used $V^{\Vert}$ to denote the quark potential with
respect to the parallel case. The constants $\mathcal{C}\left(u_{0}\right)$
and $\mathcal{C}^{\Vert}\left(u_{0}\right)$ depending on $u_{0}$
have to be calculated numerically and their behaviors with respect
to $u_{0}$ are given in Figure \ref{fig:4} \footnote{One may find a tail in the behavior of $\mathcal{C}\left(u_{0}\right)$
and $\mathcal{C}^{\Vert,\perp}\left(u_{0}\right)$ at $u_{0}\rightarrow u_{KK}$
in Figure \ref{fig:4} which seemingly implies they are non-monotonic
functions of $u_{0}$. The reason is that the formulas of the quark
potential becomes divergent at $u_{0}\rightarrow u_{KK}$, which is
recognized as a IR divergence in the dual theory. As an effective
description, we can introduce an IR cutoff $\varepsilon$ by $u_{0}=u_{KK}-\varepsilon$
to remove the divergence (also in the numerical calculation), so that
$\mathcal{C}\left(u_{0}\right)$ and $\mathcal{C}^{\Vert,\perp}\left(u_{0}\right)$
could become monotonic functions of $u_{0}$ above the cutoff.}. 

\subsubsection*{Perpendicular to the D7-branes}

The second situation is that the open string extends vertically to
the D7-branes as it is illustrated in Figure \ref{fig:3}. In this
case we take the static gauge as $\tau=t\in\left[0,\mathcal{T}\right],\sigma=y\in\left[-\frac{R}{2},\frac{R}{2}\right],u=u\left(y\right)$,
then the Nambu-Goto action is given as,

\begin{align}
S_{NG}^{\bot} & =-\frac{\mathcal{T}}{2\pi\alpha^{\prime}}\int dx\sqrt{-g_{00}\left[g_{yy}+g_{uu}u^{\prime}\left(y\right)^{2}\right]}\nonumber \\
 & =-\frac{L^{2}\mathcal{T}}{2\pi\alpha^{\prime}}\int\frac{1}{u^{2}}\sqrt{\mathcal{H}+\frac{u^{\prime2}}{\mathcal{F}}}dy,\label{eq:38}
\end{align}
where the derivatives `` $^{\prime}$ '' are with respect to $y$.
And the associated Hamiltonian to (\ref{eq:38}) reduces to a constant
which is,

\begin{equation}
\frac{\mathcal{H}}{u^{2}\sqrt{\mathcal{H}+\frac{u^{\prime2}}{\mathcal{F}}}}=\frac{\mathcal{H}_{0}}{u_{0}^{2}},
\end{equation}
or equivalently

\begin{equation}
\frac{du}{dy}=\frac{\sqrt{\mathcal{F}\mathcal{H}}\sqrt{\mathcal{H}u_{0}^{4}-\mathcal{H}_{0}^{2}u^{4}}}{\mathcal{H}_{0}u^{2}}.\label{eq:40}
\end{equation}
As the analysis in the parallel case, the quark-antiquark potential
$V^{\bot}$ can be obtained by plugging (\ref{eq:40}) into (\ref{eq:38})
then subtracting (\ref{eq:31}). The final result is given as,

\begin{align}
V^{\bot} & =\frac{L^{2}\mathcal{T}}{2\pi\alpha^{\prime}}\int_{u_{0}}^{0}\frac{du}{u^{2}\sqrt{\mathcal{F}}}\left(\frac{u_{0}^{2}\sqrt{\mathcal{H}}}{\sqrt{\mathcal{H}u_{0}^{4}-\mathcal{H}_{0}^{2}u^{4}}}-1\right)\nonumber \\
 & =\lambda^{1/2}M_{KK}\mathcal{C}\left(u_{0}\right)+\frac{a^{2}}{M_{KK}}\mathcal{C}^{\bot}\left(u_{0}\right)+\mathcal{O}\left(a^{4}\right),
\end{align}
where the constant $\mathcal{C}\left(u_{0}\right)$ and $\mathcal{C}^{\bot}\left(u_{0}\right)$
as functions of $u_{0}$ are given in Figure \ref{fig:4}.

As we have required that the size of the compactified direction is
sufficiently small $\delta z\rightarrow0$, so that the functions
$\mathcal{F},\mathcal{B},\phi$ in (\ref{eq:6}) are analytical. Thus
the quark tension can be evaluated analytically. In the large $R$
limit, to minimize its energy, the fundamental string trends to stretch
as much as possible over $u=u_{KK}$. According to (\ref{eq:38}),
its effective tension would be proportional to $\sqrt{-g_{00}g_{xx,yy}}$
and the fundamental string would move approximately vertically up
to UV cutoff around the extrema $x=\pm R/2$. Therefore in the large
$R$ limit we can obtain

\begin{equation}
V^{\Vert,\bot}\simeq\frac{R}{2\pi\alpha^{\prime}}\sqrt{-g_{00}g_{xx,yy}}\big|_{u=u_{KK}}\equiv T_{s}^{\Vert,\bot}R,
\end{equation}
which illustrates an area law of the Wilson loop. So the quark tension
is evaluated as,

\begin{align}
T_{s}^{\Vert} & =\frac{1}{2\pi\alpha^{\prime}}\sqrt{-g_{00}g_{xx}}\big|_{u=u_{KK}}\simeq\frac{\lambda^{1/2}M_{KK}^{2}}{8\pi}-\frac{\log32-2}{48\pi}\lambda^{1/2}a^{2}+\mathcal{O}\left(a^{4}\right),\nonumber \\
T_{s}^{\bot} & =\frac{1}{2\pi\alpha^{\prime}}\sqrt{-g_{00}g_{yy}}\big|_{u=u_{KK}}\simeq\frac{\lambda^{1/2}M_{KK}^{2}}{8\pi}-\frac{\log4-2}{48\pi}\lambda^{1/2}a^{2}+\mathcal{O}\left(a^{4}\right).\label{eq:43}
\end{align}
As we can see the quark tension in the parallel case decreases while
it increases in the perpendicular case. And this result is in agreement
with our numerical calculation in the limit $u_{0}\rightarrow u_{KK}$.
While it is not strictly to discuss the case that the anisotropy becomes
large, the quark tension $T_{s}^{\Vert}$ could be vanished if $a$
increases, which may imply the deconfinement.

\subsection{Entanglement entropy}

Since the entanglement entropy is expected to be a probe to characterize
the phase transition in the field theory, in this subsection let us
compute the entanglement entropy in the anisotropic background (\ref{eq:6}). 

We will take into account the region A and its complement, the region
B, as two physically disjoint spatial regions in the dual theory.
Based on the AdS/CFT correspondence \cite{key-36,key-52}, in the
dual theory the quantum entanglement entropy between region A and
B is identified to be the surface $\gamma$ stretched in the bulk
whose boundary coincides with the boundary of A. In general the classical
area of surface $\gamma$ in the correspondence of AdS\textsubscript{d+2}/CFT\textsubscript{d+1}
is given as,

\begin{equation}
S_{\gamma}=\frac{1}{4G_{N}^{\left(d+2\right)}}\int_{\gamma}d^{d}x\sqrt{g_{\mathrm{ind}}},\label{eq:44}
\end{equation}
where $G_{N}^{\left(d+2\right)}$ is the Newton constant in $d+2$
dimensional spacetime and $g_{\mathrm{ind}}$ is the induced metric
on $\gamma$. In order to represent the entanglement entropy at a
fixed time, surface $\gamma$ must be space-like. A natural generalization
of (\ref{eq:44}) to the ten-dimensional geometry in string theory
is

\begin{equation}
S_{\gamma}=\frac{1}{4G_{N}^{\left(10\right)}}\int_{\gamma}d^{8}xe^{-2\phi}\sqrt{g_{\mathrm{ind}}},\label{eq:45}
\end{equation}
where the induced metric $g_{\mathrm{ind}}$ should be given in the
string frame and the entanglement entropy can be obtained by minimizing
the action (\ref{eq:45}). As the most simple case, we consider the
``slab'' geometry of A as $\mathbb{R}\times l$, however it would
be straightforward to realize that the resultant entanglement entropy
given by (\ref{eq:45}) depends on the configuration of the slab as
we have seen in the cases of studying the Wilson loop, since the background
metric (\ref{eq:6}) is anisotropic in the $\left\{ x,y\right\} $
plane. Therefore let us proceed the calculations to obtain the entanglement
entropy in two cases: parallel and perpendicular case, which is similar
as what we have analyzed with the setup of Wilson loop in the previous
section\footnote{It also provides a parallel setup to compute the holographic entanglement
entropy in anisotropic supergravity background in \cite{key-53,key-54}. }. 
\begin{figure}
\begin{centering}
\includegraphics[scale=0.33]{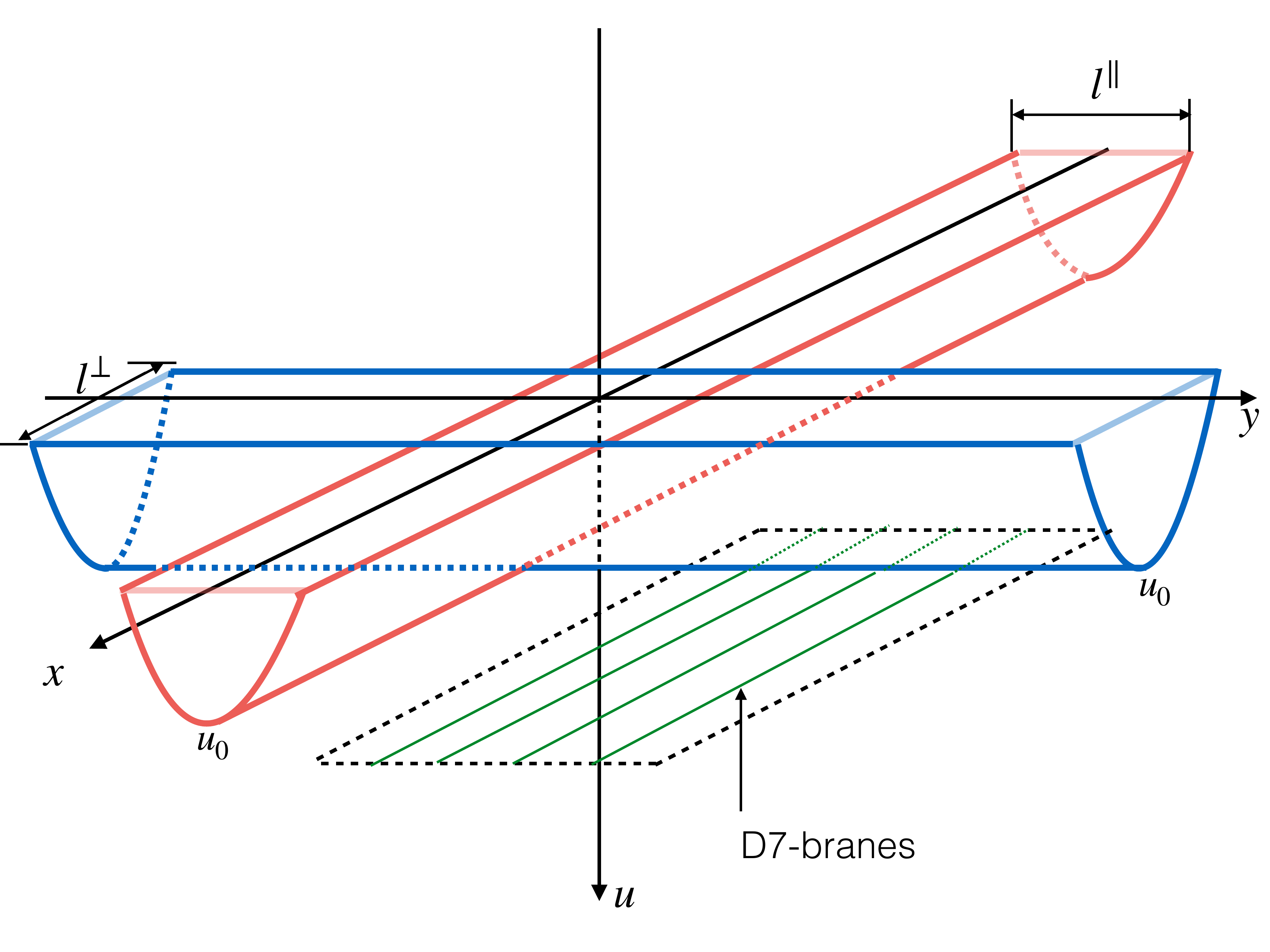}
\par\end{centering}
\caption{\label{fig:5} Surface stretched into the anisotropic bulk geometry.
The red and blue case refers to that the slab is parallel and perpendicular
to the D7-branes respectively. The D7-branes are denoted by the green
lines.}

\end{figure}

\subsubsection*{Perpendicular case}

\begin{figure}
\begin{centering}
\includegraphics[scale=0.35]{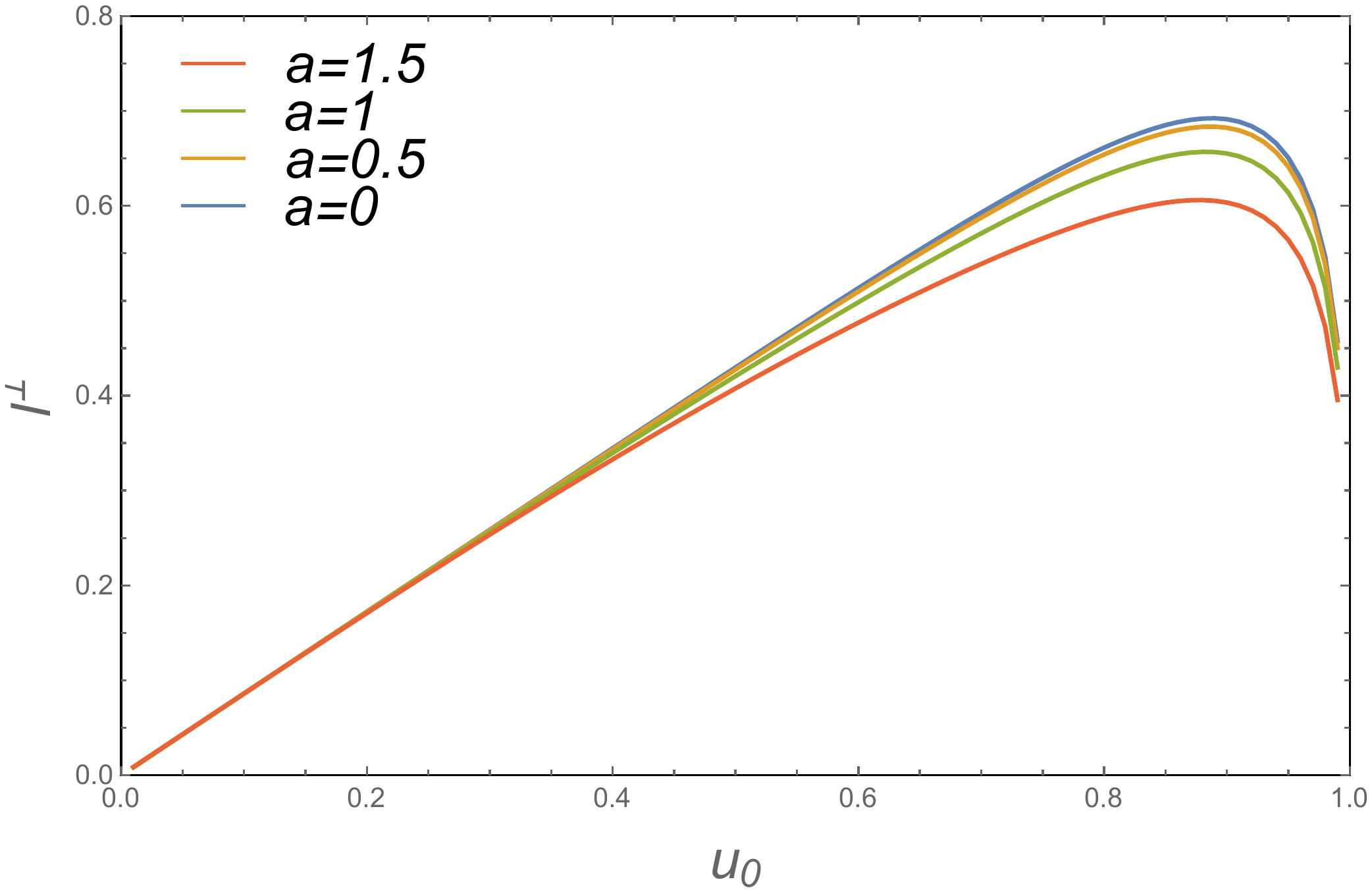}\includegraphics[scale=0.35]{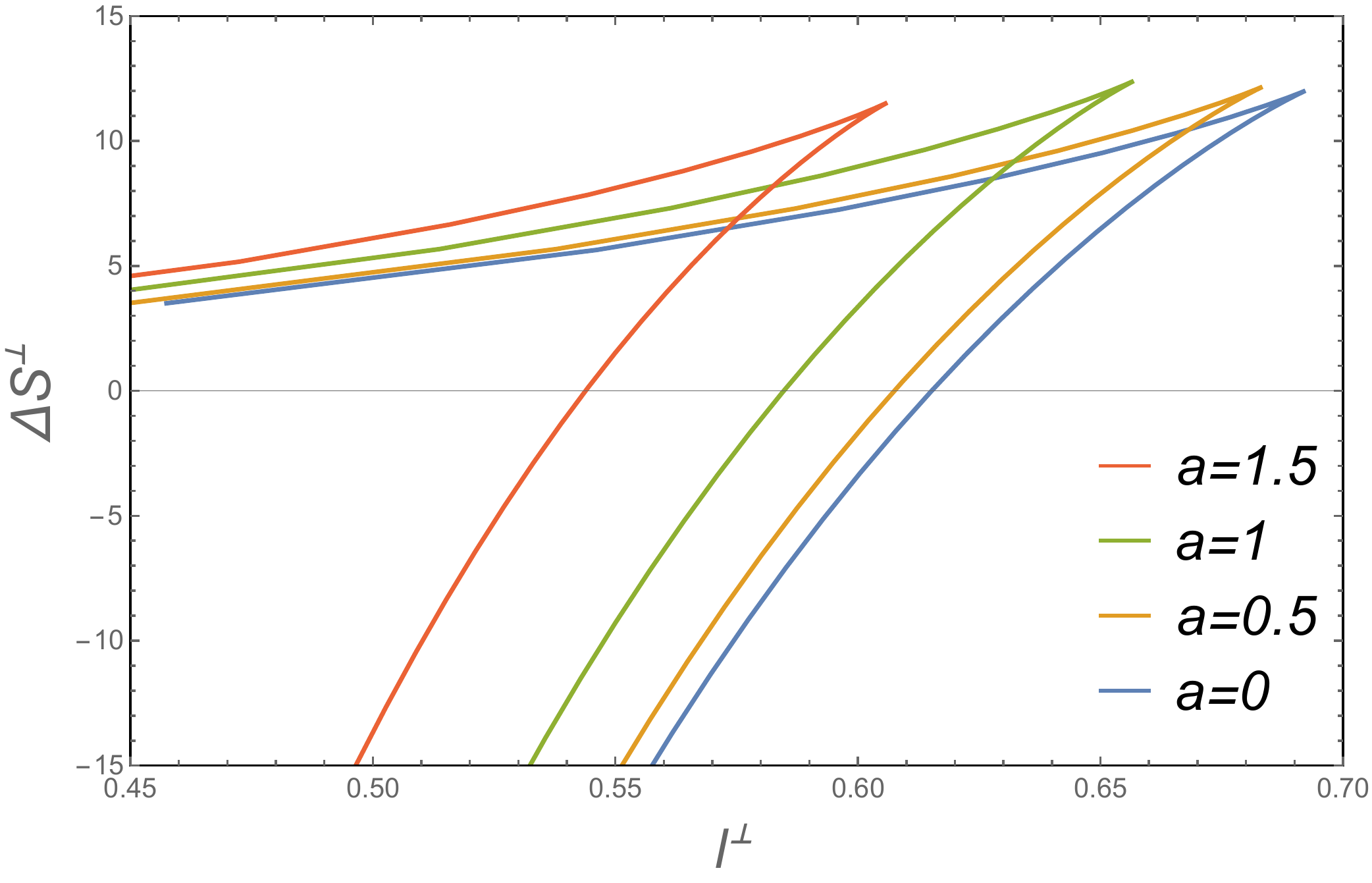}
\par\end{centering}
\begin{centering}
\includegraphics[scale=0.35]{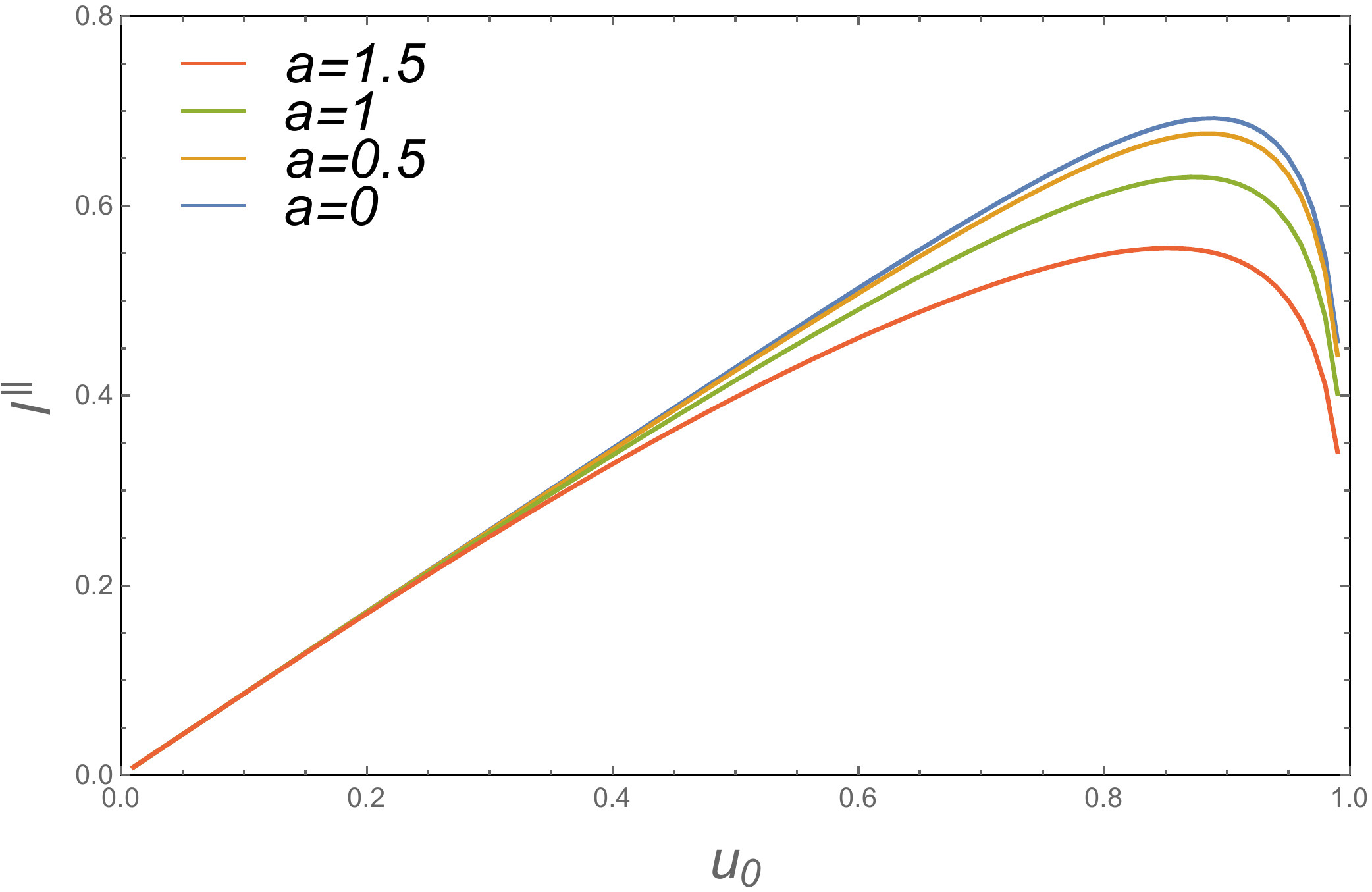}\includegraphics[scale=0.35]{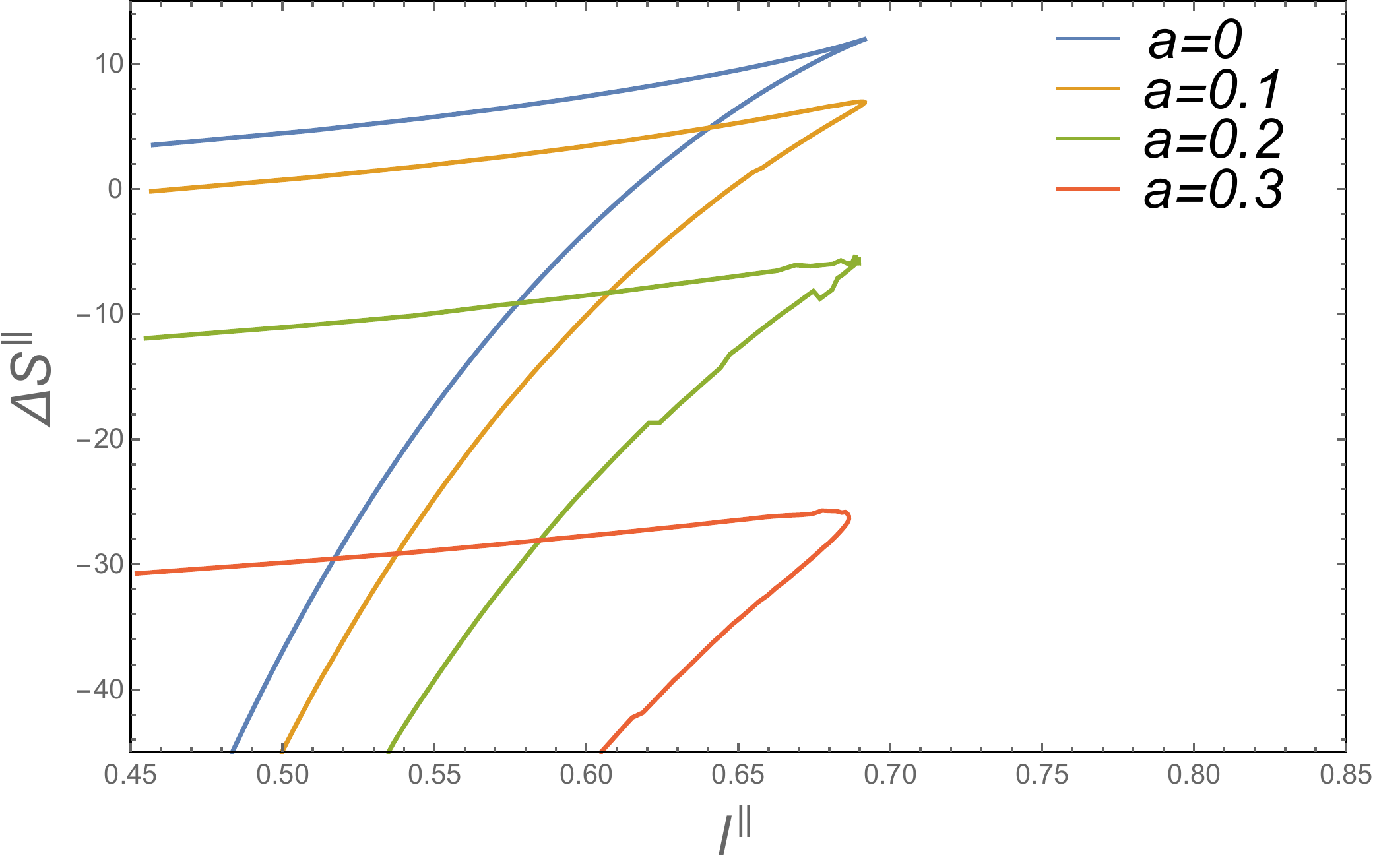}
\par\end{centering}
\caption{\label{fig:6}\textbf{ }The behavior of width and\textbf{ }entanglement
entropy in perpendicular and parallel case.\textbf{ Upper:} The relation
of $l^{\bot}$ and $u_{0}$, $\Delta S^{\bot}$ and $l^{\bot}$ for
perpendicular case. \textbf{Lower:} The relation of $l^{\Vert}$ and
$u_{0}$, $\Delta S^{\Vert}$ and $l^{\Vert}$ for parallel case.
We can see the typical swallow-tail behavior representing phase transition.}

\end{figure}

Let us first deal with the case that the region A is perpendicular
to the $N_{\mathrm{D7}}$ D7-branes as it is illustrated in Figure
\ref{fig:5}. The side of region $\gamma$ reduces to a curved line
in $\left\{ u,x\right\} $ plane, hence $u$ becomes a function of
$x$ in the induced metric. Impose (\ref{eq:6}) into (\ref{eq:45}),
after simple calculations we obtain the action of $\gamma$ as,

\begin{equation}
S^{\bot}=\frac{V}{4G}\int_{-\frac{l^{\bot}}{2}}^{\frac{l^{\bot}}{2}}dx\sqrt{h\left(u\right)}\sqrt{1+\beta u^{\prime2}},\label{eq:46}
\end{equation}
where the derivatives `` $^{\prime}$ '' are with respect to $x$,
$V$ refers to the infinity volume of the two-dimensional worldvolume
and

\begin{equation}
\alpha\left(u\right)=\frac{L^{2}}{u^{2}},\beta\left(u\right)=\frac{1}{\mathcal{F}},h\left(u\right)=e^{-4\phi}V_{\mathrm{int}}^{2}\alpha\left(u\right)^{2},V_{\mathrm{int}}=2\pi^{4}R_{3}\frac{\mathcal{Z}^{5/2}L^{6}}{u}\sqrt{\mathcal{F}\mathcal{B}}.
\end{equation}
The associated Hamiltonian to (\ref{eq:46}) must be a constant since
the Lagrangian presented in (\ref{eq:46}) is independent on $x$.
Thus it leads to,

\begin{equation}
\frac{du}{dx}=\frac{1}{\sqrt{\beta}}\sqrt{\frac{h\left(u\right)}{h\left(u_{0}\right)}-1},
\end{equation}
i.e.

\begin{equation}
l^{\bot}\left(u_{0}\right)=2\sqrt{h\left(u_{0}\right)}\int_{0}^{u_{0}}\frac{du\sqrt{\beta\left(u\right)}}{\sqrt{h\left(u\right)-h\left(u_{0}\right)}},
\end{equation}
where $l^{\bot}$ refers to the width of the region A. Obviously,
the entanglement entropy given by (\ref{eq:46}) is divergent since
it is proportional to the area of region A which therefore has to
be renormalized. The ``counterterm'' can be obtained by evaluating
the action of a surface extending along $u$. Afterwards, the finite
entanglement entropy follows the formulas as,

\begin{equation}
\frac{2G}{V}\Delta S^{\bot}=\int_{0}^{u_{0}}du\sqrt{\beta h}\left(\frac{1}{\sqrt{1-\frac{h\left(u_{0}\right)}{h\left(u\right)}}}-1\right)-\int_{u_{0}}^{u_{KK}}du\sqrt{\beta h}.
\end{equation}
By varying the value of $a$, the relation of $l^{\bot}$ and $u_{0}$,
$\Delta S^{\bot}$ and $l^{\bot}$ is illustrated numerically in Figure
\ref{fig:6} which displays the typical swallow-tail behavior with
respect to $\Delta S^{\bot}$ and $l^{\bot}$. 
 Notice that the critical value of $l^{\bot}$ , denoted by $l_{c}^{\bot}$
satisfying $\Delta S^{\bot}\left(l_{c}^{\bot}\right)=0$ decreases
when $a$ increases. Although our numerical calculation may be exactly
valid for small anisotropy, it implies the deconfinement phase transition
may trend to be vanished when $a$ increases if the entanglement entropy
does characterize the confinement as in \cite{key-36,key-37,key-38,key-39}. 

\subsubsection*{Parallel case}

Let us turn to the case that the region A is parallel to the D7-branes
as it is illustrated in Figure \ref{fig:5}. Using (\ref{eq:45}),
the action of $\gamma$ is given as, 

\begin{equation}
S^{\Vert}=\frac{V}{4G}\int_{-\frac{l^{\Vert}}{2}}^{\frac{l^{\Vert}}{2}}dx\sqrt{h\left(u\right)}\sqrt{\mathcal{H}+\beta u^{\prime2}},\label{eq:51}
\end{equation}
where the derivatives `` $^{\prime}$ '' are with respect to $y$.
Then we can obtain the relation

\begin{equation}
\frac{du}{dy}=\sqrt{\frac{h\mathcal{H}^{2}-h_{0}\mathcal{H}_{0}\mathcal{H}}{h_{0}\mathcal{H}_{0}\beta}},
\end{equation}
where $h_{0}=h\left(u_{0}\right),\mathcal{H}_{0}=\mathcal{H}\left(u_{0}\right)$
since the associated Hamiltonian is constant. Thus the the width $l^{\Vert}$
is given as,

\begin{equation}
l^{\Vert}\left(u_{0}\right)=2\sqrt{h_{0}}\int_{0}^{u_{0}}\frac{du\sqrt{\beta}}{\sqrt{h-h_{0}}}.
\end{equation}
By subtracting the divergence in (\ref{eq:51}), the finite part of
the entanglement entropy is given as,

\begin{equation}
\frac{2G}{V}\Delta S^{\Vert}=\int_{0}^{u_{0}}\sqrt{\beta H}\left(\frac{1}{\sqrt{\mathcal{H}-\frac{\mathcal{H}_{0}H_{0}}{\mathcal{H}H}}}-1\right)-\int_{u_{0}}^{u_{KK}}du\sqrt{\beta H}.
\end{equation}
And the relation of $l^{\Vert}$ and $u_{0}$, $\Delta S^{\Vert}$
and $l^{\Vert}$ is also illustrated in Figure \ref{fig:6}. While
the numerical calculation shows the typical swallow-tail behavior
with respect to $\Delta S^{\Vert}$ and $l^{\Vert}$, the associated
phase transition trends to be vanished since there would not be a
critical $l^{\Vert}$ satisfying $\Delta S^{\Vert}\left(l^{\Vert}\right)=0$
if anisotropy becomes sufficiently large. And again this conclusion
is seemingly in agreement with the analysis of the Wilson loop if
the entanglement entropy characterizes the confinement.

\subsection{Baryon vertex}

In the gauge-gravity duality, the baryon vertex is identified as a
probe D-brane wrapped on the additional dimensions denoted by the
spherical coordinates with $N_{c}$ open strings \cite{key-40}, and
it can be treated as operator to create the baryon state\footnote{In the gauge-gravity duality, the baryon state is created by quantizing
the baryonic brane somehow. While it may be a little tricky in IIB
string theory, one could review the quantization of the baryonic brane
in IIA string theory by the approach of instanton \cite{key-61} and
matrix model \cite{key-62}.}. Accordingly, in the type IIB supergravity on $\mathrm{AdS}_{5}\times S^{5}$,
the baryon vertex is a D5-brane wrapped on $S^{5}$. 

To search for the stable wrapped configuration of D5-brane in the
background (\ref{eq:6}), it would be straightforward to investigate
the condition of the force balance for the baryon vertex. To begin
with, let us decompose the metric on $\Omega_{5}$ by the coordinates
of polar angle $\eta$ and $\Omega_{4}$, then define the radius coordinate
$\xi$ as,

\begin{equation}
\frac{u_{KK}^{2}}{u^{2}}=\frac{1}{2}\left(\xi^{2}+\xi^{-2}\right).
\end{equation}
Therefore the metric (\ref{eq:6}) becomes,

\begin{equation}
ds^{2}=\frac{L^{2}}{u^{2}}\left(-dt^{2}+dx^{2}+\mathcal{H}dy^{2}+\mathcal{F}\mathcal{B}dz^{2}\right)+\frac{L^{2}}{\xi^{2}}\mathcal{Y}d\xi^{2}+L^{2}\mathcal{Z}\left(d\eta^{2}+\sin^{2}\eta d\Omega_{4}^{2}\right),
\end{equation}
where

\begin{equation}
\mathcal{Y}=\left(1-\frac{u^{4}}{u_{KK}^{4}}\right)\mathcal{F}^{-1},
\end{equation}
which implies $\xi$ and $\eta$ refers respectively to the radius
and polar angle in $\left\{ \xi,\eta\right\} $ plane. Since the baryon
vertex D5-brane extends along the directions of $\left\{ t,\eta,\Omega_{4}\right\} $,
the induced metric on a probe D5-brane is,

\begin{equation}
ds_{\mathrm{D5}}^{2}=-\frac{L^{2}}{u^{2}}dt^{2}+L^{2}\left(\mathcal{Y}\frac{\xi^{\prime2}}{\xi^{2}}+\mathcal{Z}\right)d\eta^{2}+L^{2}\mathcal{Z}\sin^{2}\eta d\Omega_{4}^{2},\label{eq:58}
\end{equation}
where the derivatives `` $^{\prime}$ '' are with respect to $\eta$.
To include the baryon potential, we turn on a single component of
the gauge field on the D5-brane as $A=A_{t}\left(\eta\right)dt$,
then the effective action for such a D5-brane is,

\begin{align}
S_{\mathrm{D5}} & =-T_{5}\int d^{6}xe^{-\phi}\sqrt{-\det\left(g_{\mathrm{D5}}+2\pi\alpha^{\prime}F\right)}-T_{5}\int A\wedge F_{5}\equiv-\frac{T_{5}V_{S^{4}}L^{6}}{\sqrt{2}u_{KK}}\int\mathcal{L}_{\mathrm{D5}}d\eta dt,\label{eq:59}
\end{align}
where $F=dA,T_{5}=\left(2\pi\right)^{-5}l_{s}^{-6}g_{s}^{-1}$ and

\begin{align}
\mathcal{L}_{\mathrm{D5}} & =\sin^{4}\eta\left[\sqrt{\left(\mathcal{Y}\xi^{\prime2}+e^{\phi/2}\xi^{2}\right)\left(1+\xi^{-4}\right)-\tilde{F}_{t\eta}^{2}}-4\tilde{A}_{t}\right],\nonumber \\
\tilde{A}_{t} & =\frac{2\sqrt{2}\pi\alpha^{\prime}u_{KK}}{L^{2}}A_{t},\tilde{F}_{t\eta}=-\partial_{\eta}\tilde{A}_{t}.\label{eq:60}
\end{align}
Defining the displacement \cite{key-44,key-55},

\begin{equation}
D\left(\eta\right)=\frac{\partial\mathcal{L}_{\mathrm{D5}}}{\partial\tilde{F}_{t\eta}}=-\frac{\sin^{4}\eta\tilde{F}_{t\eta}}{\sqrt{\left(\mathcal{Y}\xi^{\prime2}+e^{\phi/2}\xi^{2}\right)\left(1+\xi^{-4}\right)-\tilde{F}_{t\eta}^{2}}},
\end{equation}
the equation of motion for $\tilde{A}_{t}$ can be written as,
\begin{figure}[th]
\begin{centering}
\includegraphics[width=7.4cm,height=4.8cm]{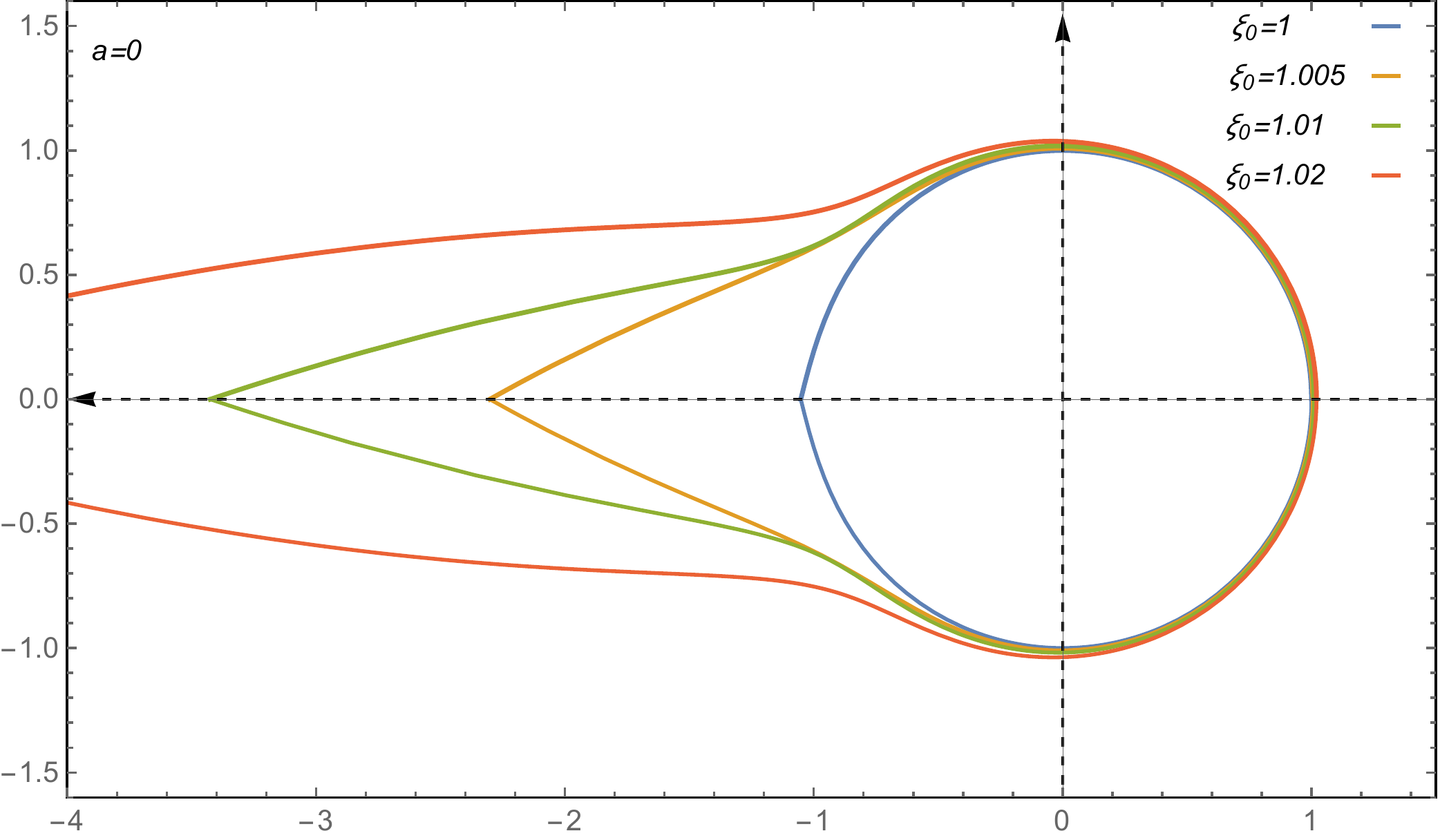}\includegraphics[width=7.4cm,height=4.8cm]{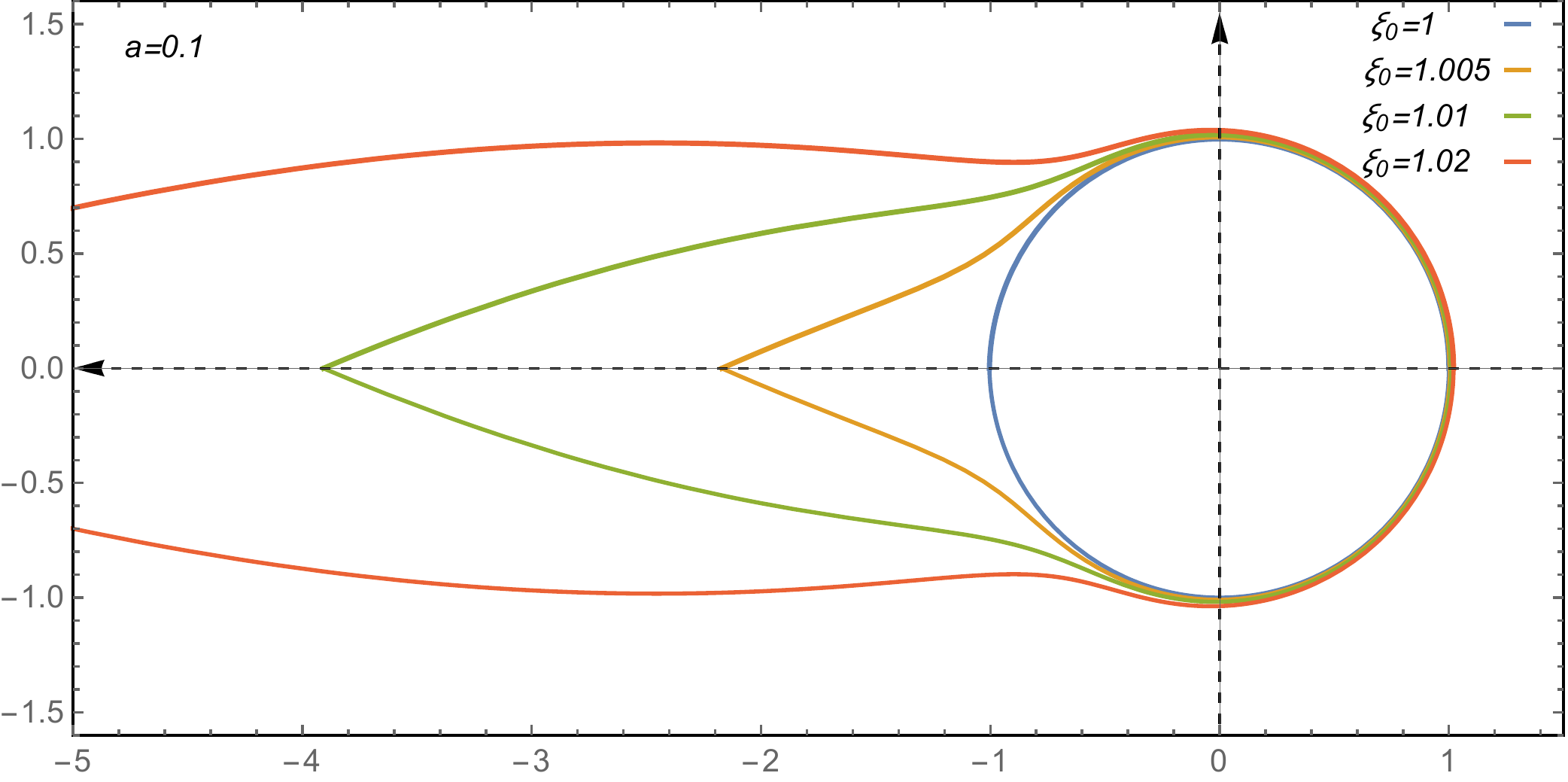}
\par\end{centering}
\begin{centering}
\includegraphics[width=7.4cm,height=4.8cm]{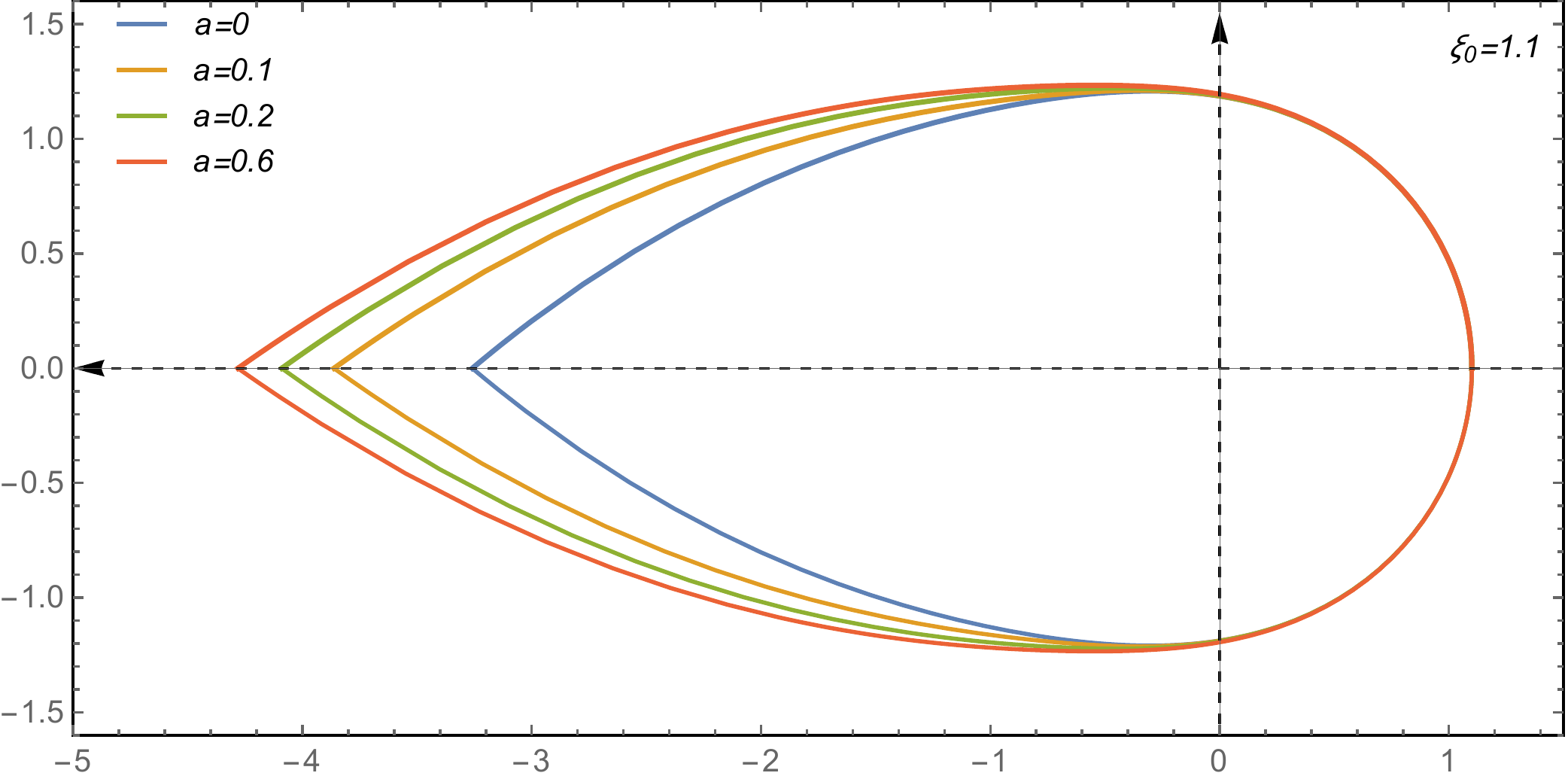}\includegraphics[width=7.4cm,height=4.8cm]{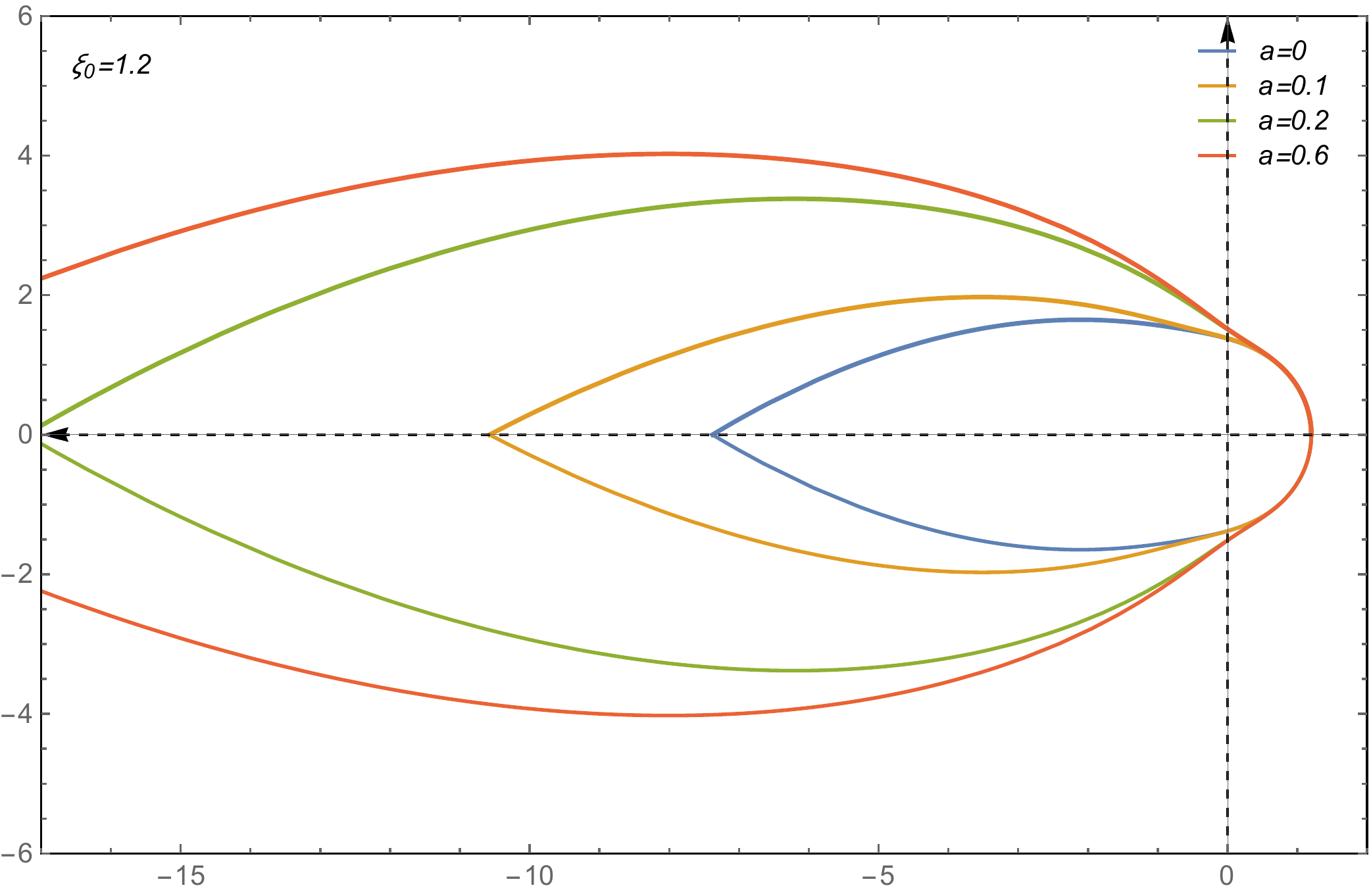}
\par\end{centering}
\caption{\label{fig:7} The configuration of the D5-brane as baryon vertex
in $\left\{ \xi,\eta\right\} $ plane. Note $\xi$ and $\eta$ is
the radius and polar angle. In the upper figures, we fix $a=0,0.1$
respectively then adjust $\xi_{0}$; in the lower figures, we fix
$\xi_{0}=1.1,1.2$ then adjust $a$. The numerical calculation always
illustrates wrapped shape of the baryon vertex.}
\end{figure}

\begin{equation}
\partial_{\eta}D\left(\eta\right)=-4\sin^{4}\eta.
\end{equation}
So the displacement is solved as,

\begin{equation}
D\left(\eta\right)=-\frac{3}{2}\eta+\frac{3}{2}\sin\eta\cos\eta+\sin^{3}\eta\cos\eta,
\end{equation}
and the associated Hamiltonian to (\ref{eq:60}) is,

\begin{align}
\mathcal{H}_{\mathrm{D5}} & =\tilde{F}_{t\eta}\frac{\partial\mathcal{L}_{\mathrm{D5}}}{\partial\tilde{F}_{t\eta}}-\mathcal{L}_{\mathrm{D5}}\nonumber \\
 & =-\sqrt{\left(\mathcal{Y}\xi^{\prime2}+e^{\phi/2}\xi^{2}\right)\left(1+\xi^{-4}\right)}\sqrt{D^{2}\left(\eta\right)+\sin^{8}\eta}.
\end{align}
Accordingly the force at the cusp $U=U_{c}$ of D5-brane is given
as,

\begin{equation}
F_{\mathrm{D5}}=-\frac{T_{5}V_{S^{4}}L^{6}}{\sqrt{2}u_{KK}}\int_{0}^{\pi}\frac{\partial\mathcal{H}_{\mathrm{D5}}}{\partial U_{c}}=N_{c}T_{F}\frac{\xi_{c}^{-4}+1}{\xi_{c}^{-4}-1}\frac{\mathcal{Y}_{c}\xi_{c}^{\prime}}{\sqrt{\mathcal{Y}_{c}\xi_{c}^{\prime2}+e^{\phi_{c}}\xi_{c}^{2}}},
\end{equation}
where $U=L^{2}/u$ is the standard radius coordinate of the bulk and
we have used the index $c$ to refer to the value of the variables
at $U=U_{c}$. For stable configuration located at $u_{c}=L^{2}/U_{c}$,
the D5-brane must satisfy the zero-force condition i.e. $F_{\mathrm{D5}}=0$
if there is no other probe brane. It implies the zero-force condition
can be achieved only if $\xi_{c}^{\prime}=0$ since $\mathcal{Y}$
is always positive in the bulk. Therefore the stable position for
the baryon vertex in this system is located at $u_{c}\rightarrow u_{KK}$
to consistently minimize its action. To confirm our analysis, we also
solve numerically the equation of motion associated to the Lagrangian
presented in (\ref{eq:60}) to obtain the embedding function of the
probe D5-brane\footnote{To obtain the equation of motion for the embedding function, we can
set $A=0$ in the Lagrangian presented in (\ref{eq:60}).}. For a stable solution, we impose the boundary condition $\xi^{\prime}\left(0\right)=0,\xi\left(0\right)=\xi_{0}$
to the equation of motion, then the numerical solution for the configuration
of the baryon vertex in $\left\{ \xi,\eta\right\} $ plane is given
in Figure \ref{fig:7}. As we can see, there always exists wrapped
solution for the baryon vertex and the configuration of $\xi_{0}=1$
is nearly independent on the variation of $a$. Thus the stable position
for the D5-brane is indeed $u_{c}\rightarrow u_{KK}$ if the baryon
vertex is the only probe brane.

To close this subsection, let us evaluate the baryon mass in this
holographic system. Since the stable baryon vertex must be located
at $u_{KK}$, its mass can be obtained by evaluate its onshell action
(\ref{eq:59}) by setting $A=0$. Thus the Euclidean action of the
D5-brane $S_{\mathrm{D5}}^{E}$ and baryon mass $m_{B}$ are identified
via holography as,

\begin{equation}
S_{\mathrm{D5}}^{E}=T_{5}\int d^{6}xe^{-\phi}\sqrt{-\det g_{\mathrm{D5}}}\big|_{u=u_{KK}}=m_{B}\int dt.\label{eq:66}
\end{equation}
Plugging the metric in (\ref{eq:58}) and the relation of $M_{KK},u_{KK}$
in (\ref{eq:9}) into (\ref{eq:66}), the baryon mass is evaluated
as,

\begin{equation}
m_{B}=\frac{\lambda^{1/2}}{64\pi^{2}}M_{KK}N_{c}\left[1+\frac{\log16-1}{6}\frac{a^{2}}{M_{KK}^{2}}+\mathcal{O}\left(a^{4}\right)\right],
\end{equation}
which is enhanced in the presence of the Chern-Simons term represented
by $a$. As we have specified in the Section 2, the dual theory is
a topological massive theory due to the presence of the Chern-Simons
term, so it would be easy to understand that baryon would also become
topologically massive once the fundamental fermion is introduced.
In addition, as the baryon mass is proportional to the worldvolume
of the D5-brane (\ref{eq:66}), we find Figure \ref{fig:7} also illustrates
the worldvolume is increased by the anisotropy which implies the increase
of the baryon mass by the anisotropy.

\section{Embedding of the D7-branes and the vacuum structure}

In this section, we continue the holographic setup by introducing
the various D7-branes to identify the dual field theory as QCD\textsubscript{3}
in large $N_{c}$ limit. We first address the D7-branes as the flavor
degrees of freedom then take into account another D7-brane in which
the low-energy effective theory on its worldvolume is expected to
be pure Chern-Simons theory. And we will see, the topological property
in the dual theory can be studied by analyzing the configuration and
orientation of these distinct D7-branes.

\subsection{The flavor brane}

As most works about the gauge-gravity duality, the fundamental matter
can be added to the D3-brane background by embedding flavor D7-branes
as probe. In our setup, we add $N_{f}$ coincident copies of D7-branes
as probes to the D3-brane background (\ref{eq:6}), transverse to
the compactified $z$ direction, spanning the $\mathbb{R}^{1,2}$
denoted by $\left\{ t,x,y\right\} $, the holographic direction denoted
by $u$ and four of the five directions in $\Omega_{5}$ as \cite{key-45,key-46}.
The D-brane configuration including various D7-branes is illustrated
in Table \ref{tab:3} where we have decomposed the directions of $\Omega_{5}$
as $\Omega_{4}$ and $w$. 
\begin{table}
\begin{centering}
\begin{tabular}{|c|c|c|c|c|c|c|c|}
\hline 
Bubble background & $t$ & $x$ & $y$ & $\left(z\right)$ & $u$ & $\Omega_{4}$ & $w$\tabularnewline
\hline 
\hline 
$N_{c}$ D3-branes & - & - & - & - &  &  & \tabularnewline
\hline 
$N_{\mathrm{D7}}$ D7-branes & - & - &  & - &  & - & -\tabularnewline
\hline 
$N_{f}$ D7-branes & - & - & - &  & - & - & \tabularnewline
\hline 
CS D7-branes & - & - & - &  &  & - & -\tabularnewline
\hline 
\end{tabular}
\par\end{centering}
\caption{\label{tab:3} The D-brane configuration including various D7-branes.}

\end{table}
 The leftover direction $w$ is transverse to both the $N_{c}$ color
D3-branes and $N_{f}$ flavor D7-branes, so a bare mass for the flavors
can be introduced by imposing a separation between color and flavor
branes along $w$ direction at the UV boundary which breaks the parity
in QCD\textsubscript{3}. In the D3-D7 approach, the $w$ direction
corresponding to the scalar in the D7-brane worldvolume couples to
the mass operator of the fermions. And according to the gauge-gravity
duality, the profile along the transverse direction of the flavor
branes corresponds to the meson operator $\bar{\psi}\psi$ in the
dual theory.

\begin{figure}
\begin{centering}
\includegraphics[scale=0.5]{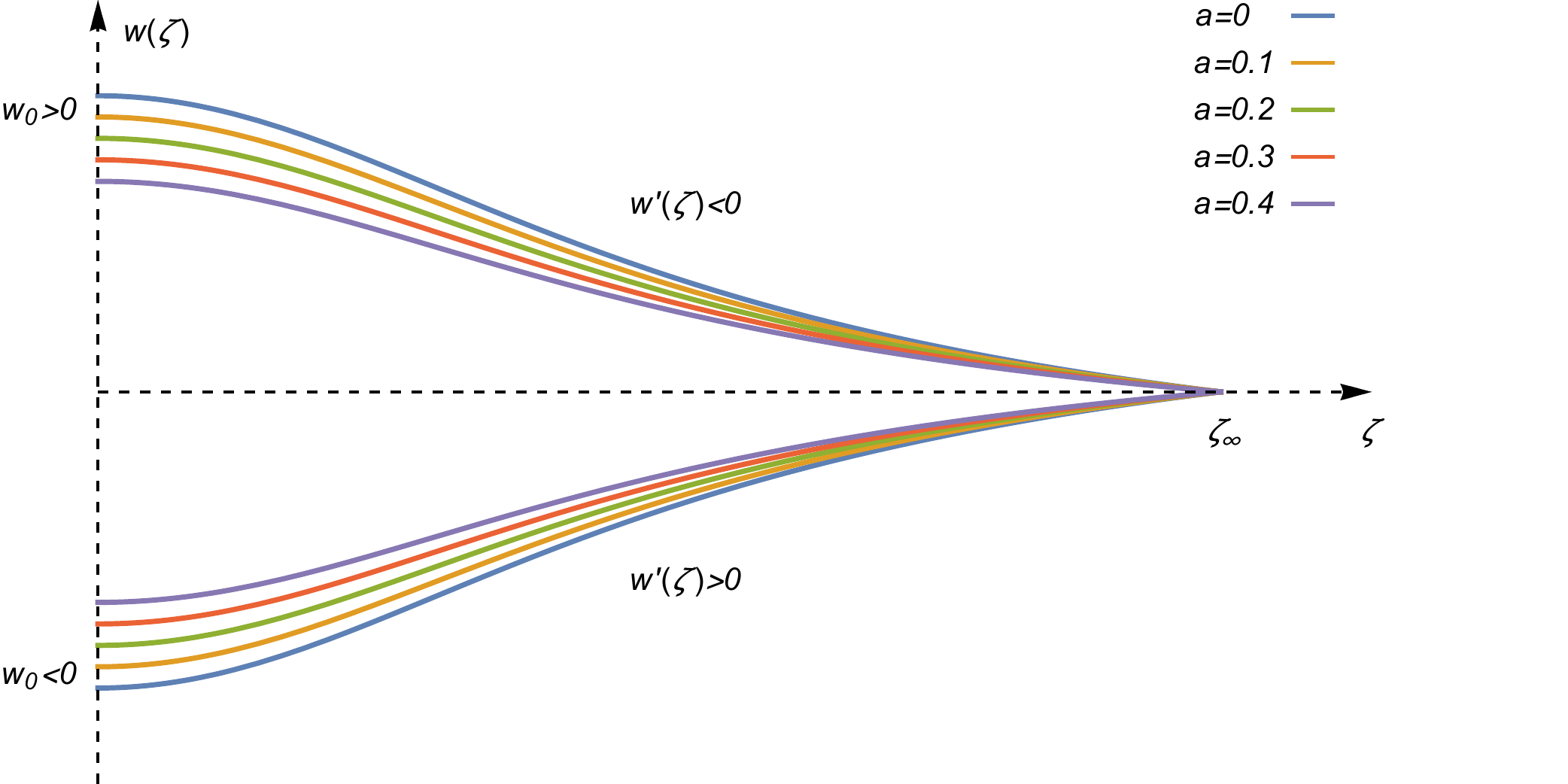}
\par\end{centering}
\begin{centering}
\includegraphics[scale=0.5]{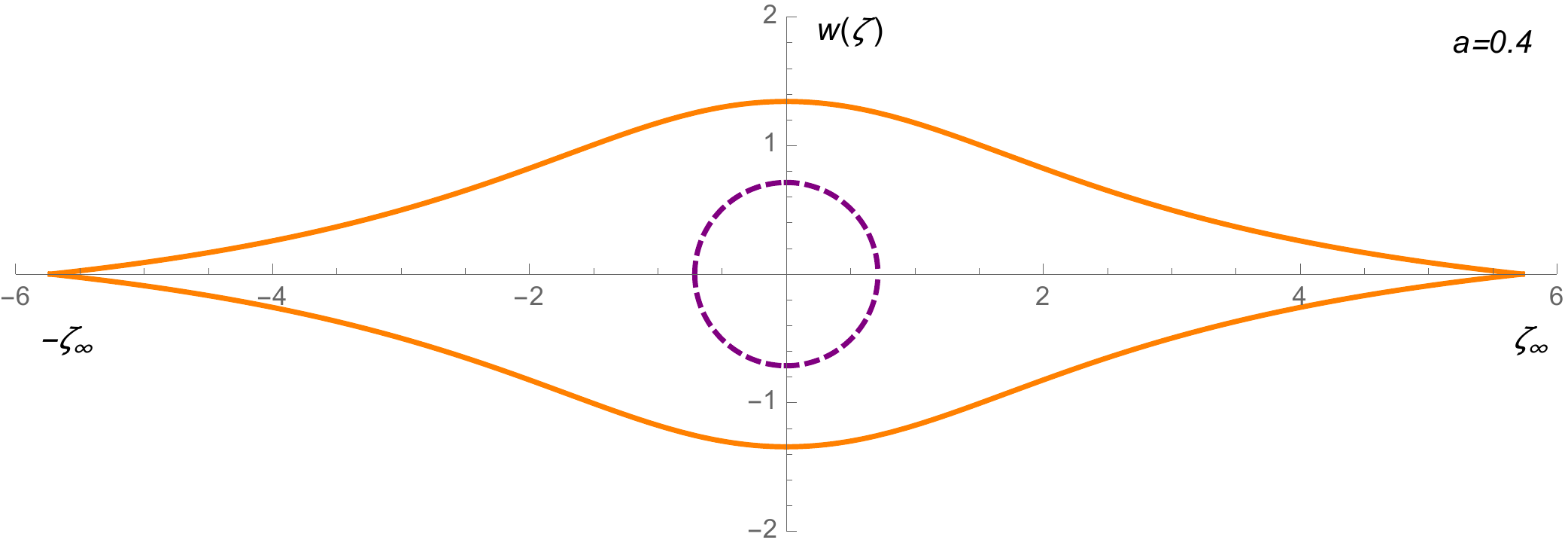}
\par\end{centering}
\caption{\label{fig:8} The two parity-related minimal embeddings of the flavor
branes in the massless case with various $a$. The parameter is chosen
as $L^{2}/u_{KK}=1,\zeta_{\infty}=5.73$. \textbf{Upper:} the dependence
on $a$ of the flavor embeddings. \textbf{Lower:} the full configuration
of the parity-related flavor branes with a fixed $a$. The purple
dashed line represents the position $u=u_{KK}$.}

\end{figure}

\begin{figure}
\begin{centering}
\includegraphics[scale=0.5]{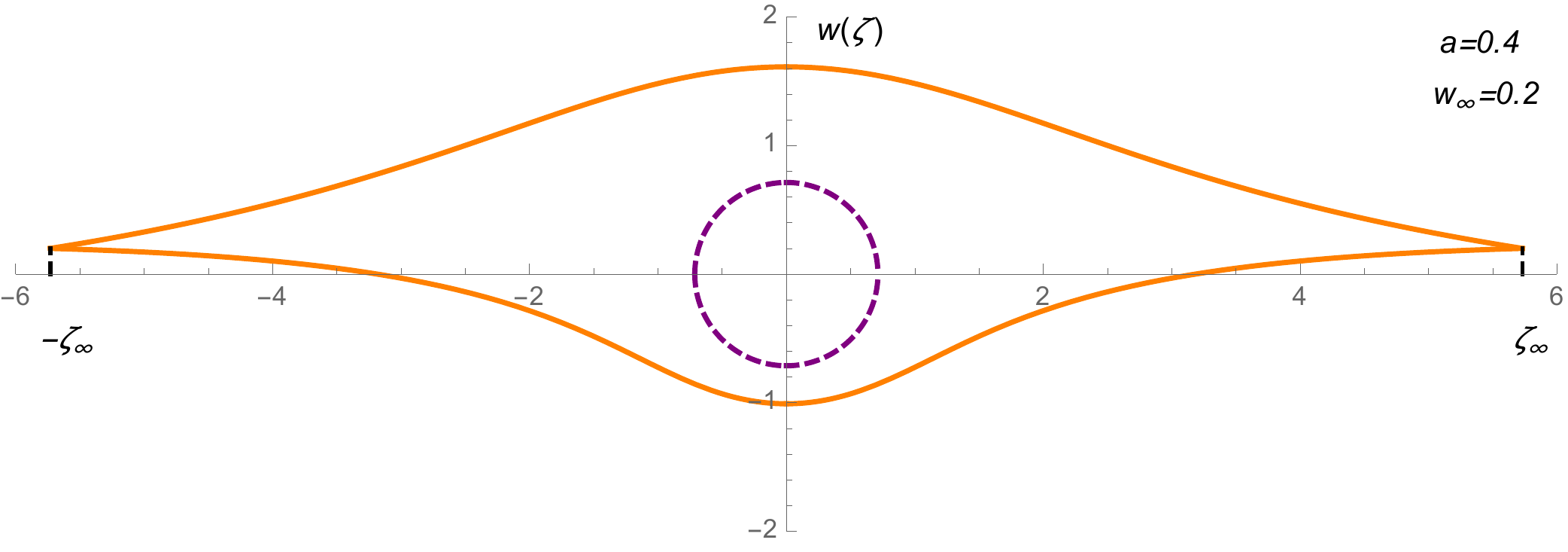}
\par\end{centering}
\begin{centering}
\includegraphics[scale=0.5]{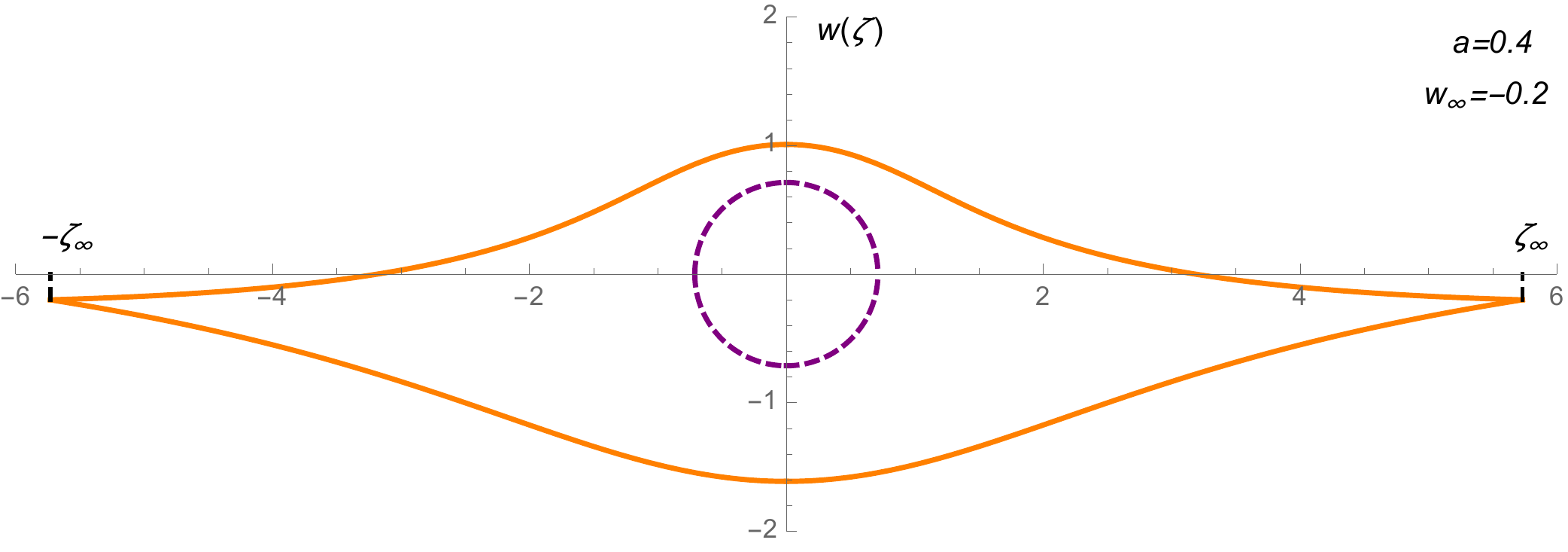}
\par\end{centering}
\caption{\label{fig:9} The massive case of the embedding flavor branes with
fixed $a$. \textbf{Upper:} the positive mass case with $w_{\infty}=0.2$.
\textbf{Lower:} the negative mass case with $w_{\infty}=-0.2$.}

\end{figure}

Then let us investigate the embedding of the flavor branes with their
effective action. To specify the embedding of the flavor branes with
respect to the transverse direction $w$, we first define a new radial
coordinate $\rho$ as,

\begin{equation}
\frac{du^{2}}{u^{2}\mathcal{F}}=\frac{\mathcal{Z}}{\rho^{2}}d\rho^{2}.
\end{equation}
Up to order of $\mathcal{O}\left(a^{2}\right)$, the relation of $u$
and $\rho$ can be solved as,

\begin{align}
u\left(\rho\right)= & u_{0}\left(\rho\right)+a^{2}u_{2}\left(\rho\right),\nonumber \\
u_{0}\left(\rho\right)= & \frac{2L^{2}u_{KK}^{2}\rho}{\sqrt{L^{8}+4u_{KK}^{4}\rho^{4}}},\nonumber \\
u_{2}\left(\rho\right)= & -\frac{L^{2}u_{KK}^{4}\rho}{24\left(L^{8}+4u_{KK}^{4}\rho^{4}\right)^{3/2}}\bigg[4L^{4}u_{KK}^{2}\rho^{2}+2L^{8}\left(\log32-1\right)\nonumber \\
 & -5\left(L^{8}+4u_{KK}^{4}\rho^{4}\right)\log\left(\frac{L^{8}+4u_{KK}^{2}\rho^{2}L^{4}+4u_{KK}^{4}\rho^{4}}{L^{8}+4u_{KK}^{4}\rho^{4}}\right)\bigg].\label{eq:69}
\end{align}
We note that the ambiguity in this relation can be omitted by choosing
$\rho^{2}\geq L^{4}/\left(u_{KK}^{2}2\right)$ for $a=0$. In this
coordinate, the background metric (\ref{eq:6}) can be written as,

\begin{equation}
ds^{2}=\frac{L^{2}}{u^{2}}\left(-dt^{2}+dx^{2}+\mathcal{H}dy^{2}+\mathcal{F}\mathcal{B}dz^{2}\right)+\frac{L^{2}\mathcal{Z}}{\rho^{2}}\left(d\rho^{2}+\rho^{2}d\Omega_{5}^{2}\right),\label{eq:70}
\end{equation}
where $u=u\left(\rho\right)$. Afterwards we impose the coordinate
transformation $\zeta=\rho\cos\Theta,w=\rho\sin\Theta$ i.e. $\rho^{2}=w^{2}+\zeta^{2}$
where $\Theta$ is one angular coordinate in $\Omega_{5}$, then the
metric (\ref{eq:70}) takes the final form as,

\begin{equation}
ds^{2}=\frac{L^{2}}{u^{2}}\left(-dt^{2}+dx^{2}+\mathcal{H}dy^{2}+\mathcal{F}\mathcal{B}dz^{2}\right)+\frac{L^{2}\mathcal{Z}}{\rho^{2}}\left(d\zeta^{2}+\zeta^{2}d\Omega_{4}^{2}+dw^{2}\right).
\end{equation}
Since the flavor D7-branes extend along $\left\{ t,x,y,u,\Omega_{4}\right\} $,
the induced metric on the flavor branes is obtained as,

\begin{equation}
ds_{\mathrm{D7}}^{2}=\frac{L^{2}}{u^{2}}\left(-dt^{2}+dx^{2}+\mathcal{H}dy^{2}\right)+\frac{L^{2}\mathcal{Z}}{\rho^{2}}\left[\left(w^{\prime2}+1\right)d\zeta^{2}+\zeta^{2}d\Omega_{4}^{2}\right],\label{eq:72}
\end{equation}
where $w=w\left(\zeta\right)$ and the derivatives `` $^{\prime}$
'' are with respect to $\zeta$. So the action for a single flavor
D7-brane is,

\begin{align}
S_{\mathrm{D7}} & =-T_{7}\int d^{8}xe^{-\phi}\sqrt{-g_{\mathrm{D7}}}=-T_{7}V_{3}V_{S^{4}}L^{8}\int d\zeta\mathcal{L},\nonumber \\
\mathcal{L} & =\frac{e^{-\phi/4}\zeta^{4}\sqrt{1+w^{\prime2}}}{u^{3}\rho^{5}}.\label{eq:73}
\end{align}
The behavior of the embedding function of the flavor D7-branes can
be obtained by solving the associated equation of motion to the action
in (\ref{eq:73}), which is,

\begin{equation}
\frac{\partial}{\partial\zeta}\left[\frac{e^{-\phi/4}\zeta^{4}w^{\prime}}{u\left(\rho\right)^{3}\rho^{5}\sqrt{1+w^{\prime2}}}\right]-\frac{\partial}{\partial w}\left[\frac{e^{-\phi/4}\zeta^{4}\sqrt{1+w^{\prime2}}}{u\left(\rho\right)^{3}\rho^{5}}\right]=0.\label{eq:74}
\end{equation}
Since a parity transformation acts as $w\left(\zeta\right)\rightarrow-w\left(\zeta\right)$,
for the massless case, we have to impose the boundary condition

\begin{equation}
w^{\prime}\left(\zeta\right)\bigg|_{\zeta=0}=0,w\left(\zeta\right)\bigg|_{\zeta=\infty}=0.
\end{equation}
Keeping these in hand, we numerically solve the equation of motion
(\ref{eq:74}) with various $a$ in the region $\zeta>0$ which is
illustrated in Figure \ref{fig:8}. 
 The numerical calculation shows a very small shift with respect to
the dependence on $a$ (we have enlarged the shift in the figure)
and this behavior is opposite to the approach of the D3-D(-1) background
\cite{key-45}. We note here the two branches of the flavor branes
trend to become coincident if $a$ increases. The full configuration
of the embedding D7-branes is given in Figure \ref{fig:8} and we
also calculate the massive case by setting the boundary condition
$w_{\infty}\neq0$ in Figure \ref{fig:9}. 
 Note that the number of the node in the embedding function refers
to the excitation of the D-brane configuration. For a vacuum configuration,
we only take the non-node solution as it is shown in Figure \ref{fig:8}
and Figure \ref{fig:9} which means equivalently the node is located
at $\zeta=\infty$.

\subsection{The Chern-Simons brane}

Due to the presence of the axion field $\chi$ in the bulk, the D3-D7
brane background (\ref{eq:6}) corresponds to the QCD\textsubscript{3}
with a Chern-Simons term in holography. Thus the present $N_{\mathrm{D7}}$
D7-branes should contribute to some vacuum properties of the dual
theory. However, once we evaluate the embedding function of such a
D7-brane as probe, its equation of motion implies the only solution
for $u=u_{KK}$ and its onshell action automatically vanishes. To
include the set-up of a pure Chern-Simons theory, we follow the discussion
in the D3-brane approach \cite{key-46}, to introduce $k_{b}$ coincident
Chern-Simons (CS) as probe D7-branes where the configuration is given
in Table \ref{tab:3}. At very low energies, all other excitations
on the Chern-Simons branes decouple and only a Wess-Zumino term is
left as,

\begin{equation}
S_{WZ}=\frac{1}{2\left(2\pi\right)^{5}l_{s}^{4}}\int C_{4}\wedge\mathrm{Tr}\left(F\wedge F\right)=-\frac{1}{2\left(2\pi\right)^{5}l_{s}^{4}}\int_{S^{5}}F_{5}\int_{\mathbb{R}^{2+1}}\omega_{3}=-\frac{N_{c}}{4\pi}\int_{\mathbb{R}^{2+1}}\omega_{3}.
\end{equation}
Therefore we can see the gauge-gravity duality in this setup reduces
to the well-known level/rank duality $SU\left(N_{c}\right)_{k_{b}}\leftrightarrow U\left(k_{b}\right)_{-N_{c}}$
in quantum field theory (QFT) expectations precisely \cite{key-45,key-46,key-56}.

\begin{figure}
\begin{centering}
\includegraphics[scale=0.36]{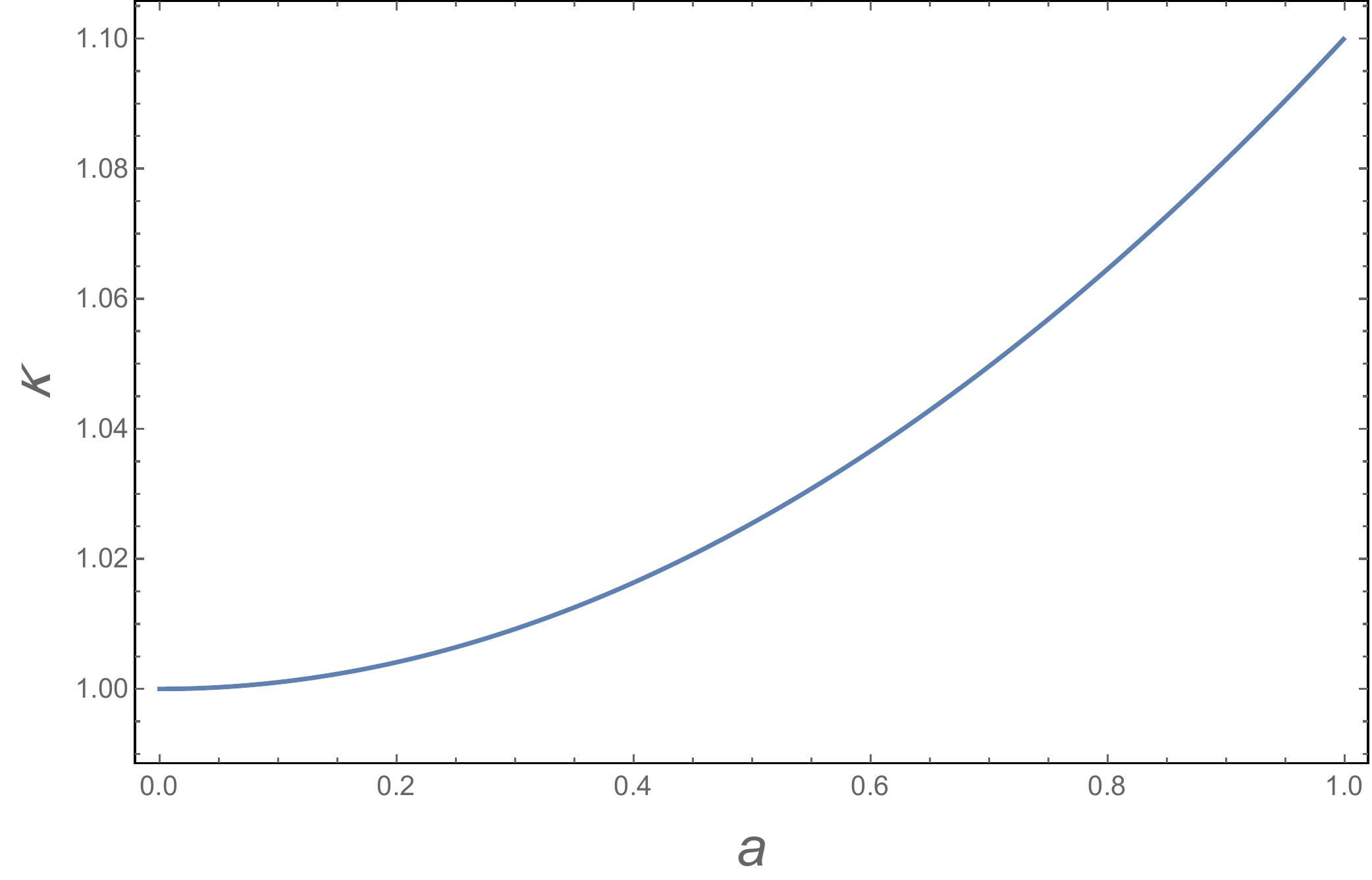}\includegraphics[scale=0.36]{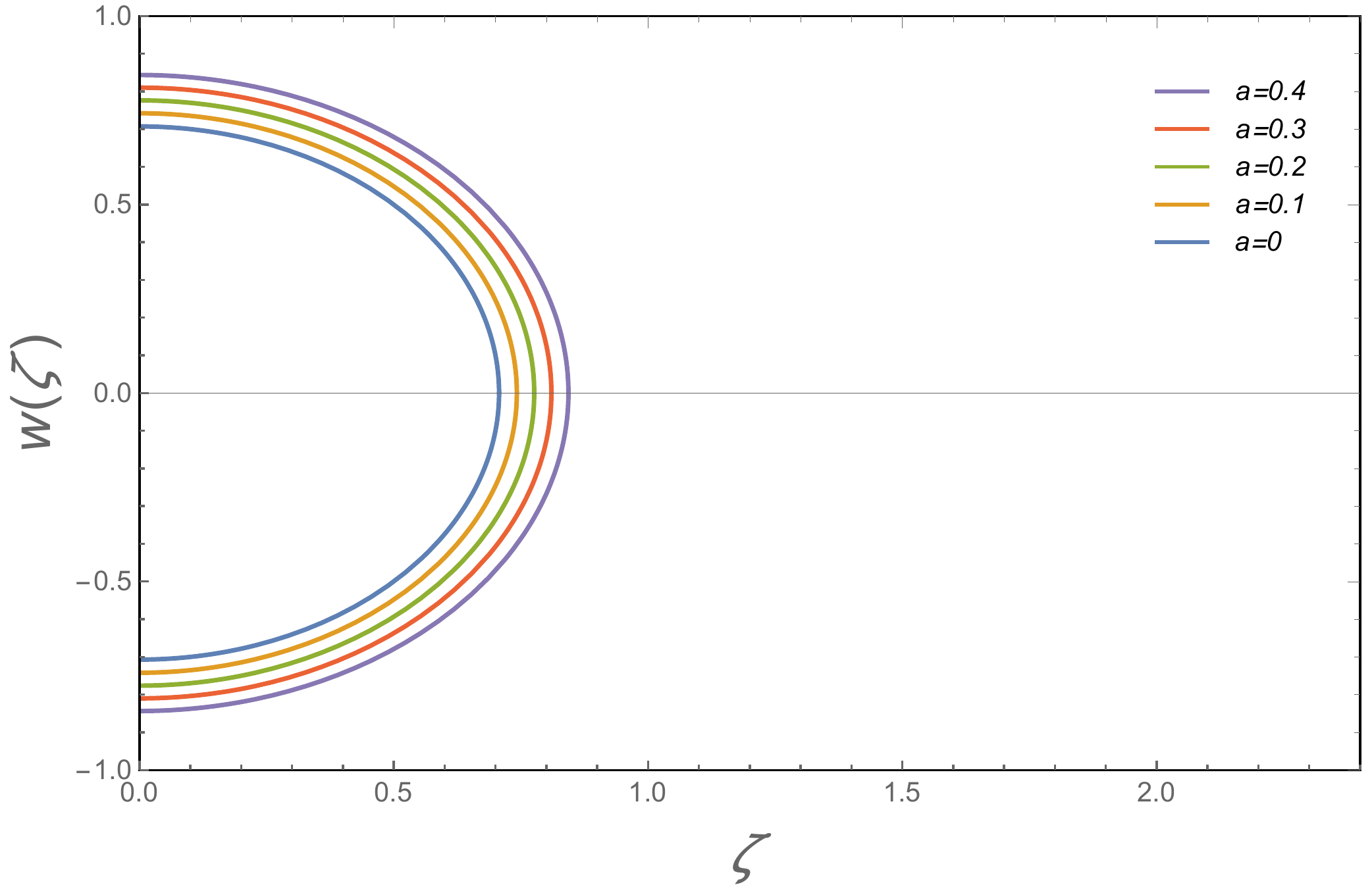}
\par\end{centering}
\caption{\label{fig:10} \textbf{Left:}The relation of $\kappa$ and $a$.
\textbf{Right: }The embedding of the Chern-Simons brane on $\left\{ w,\zeta\right\} $
plane.}

\end{figure}

To obtain the embedding of the Chern-Simons brane, we need to evaluate
the equation of motion by its effective action. Recall the induced
metric on a D7-brane (\ref{eq:72}), the action for Chern-Simons brane
takes the same formula as given in (\ref{eq:73}), thus its associated
equation of motion is given in (\ref{eq:74}) while $\rho$ must be
a constant for a Chern-Simons brane. Accordingly, we impose the following
ansatz

\begin{equation}
w\left(\zeta\right)=\sqrt{\frac{L^{4}}{2u_{KK}^{2}}\kappa-\zeta^{2}},
\end{equation}
to the equation of motion of $w\left(\zeta\right)$ which reduces
to a constraint for $\kappa$ and $a$ as,

\begin{align}
0= & -48\left(\kappa^{5}+\kappa^{4}-\kappa-1\right)+a^{2}\bigg[-6+30\log2+\kappa\big(12-16\kappa+12\kappa^{2}-18\kappa^{3}\nonumber \\
 & +\left(6+2\kappa+2\kappa^{2}\right)\log32\big)+12\left(\kappa^{5}+\kappa^{4}-\kappa-1\right)\log\left(1+\frac{2\kappa}{1+\kappa^{2}}\right)\bigg],
\end{align}
where we have set $L^{2}/u_{KK}=1$. And this constraint can be solved
numerically as it is illustrated in Figure \ref{fig:10}. 
 The numerical calculation also illustrates a very small shift with
respect to the dependence on $a$ for the embedding of the Chern-Simons
brane. Then the ground energy of the Chern-Simons brane can be evaluated
by its action as

\begin{equation}
S_{\mathrm{D7}}^{\mathrm{CS}}=-T_{7}\int d^{8}xe^{-\phi}\sqrt{-g_{\mathrm{D7}}}=-T_{7}V_{3}V_{S^{5}}L^{8}e^{-\frac{\phi}{4}}u^{-3}.
\end{equation}
The action is minimized at $u=u_{KK}$, so we can obtain the ground
energy density $E_{\mathrm{CS}}$ of the Chern-Simons brane is obtained
by

\begin{equation}
E_{\mathrm{CS}}=-\frac{S_{\mathrm{D7}}^{\mathrm{CS}}}{V_{3}}=T_{7}V_{S^{5}}L^{8}u_{KK}^{-3}\left(1+\frac{\log2}{16}a^{2}u_{KK}^{2}\right)+\mathcal{O}\left(a^{4}\right).\label{eq:80}
\end{equation}

\section{The phase diagram involving the massive flavors}

In the previous sections, we have evaluated the embedding of the Chern-Simons
and flavor branes with massless boundary condition. Here we are going
to study the configurations having both Chern-Simons and flavor branes
with massive boundary condition since the vacuum of QCD\textsubscript{3}
with Chern-Simons term would include both of them in general. Then
the phase diagram would be obtained by evaluating the energies of
these D7-branes.

\subsection{The energy of massive embedding flavor brane}

In order to obtain the energy of the flavor branes with massive boundary
condition, let us take a look at the asymptotic behavior of the embedding
functions which satisfies the equation of motion (\ref{eq:74}), although
the embedding configuration of the flavor branes has been illustrated
in Figure \ref{fig:9}. Since the flavor mass corresponds to the boundary
value of $w\left(\zeta\right)$, we can find the asymptotics of (\ref{eq:74})
at $\zeta\rightarrow\infty$ ($\rho\rightarrow\zeta$) as,

\begin{equation}
\frac{d}{d\zeta}\left(\zeta^{2}w^{\prime}\right)=-2w,
\end{equation}
where the relation (\ref{eq:69}) and boundary behavior of $\phi$
have been imposed. For massless case, the general form of the asymptotics
at large $\zeta$ for $w\left(\zeta\right)$ would be,

\begin{equation}
w\left(\zeta\right)=\pm\sqrt{\frac{\mu^{3}}{\zeta}}\sin\left(\frac{\sqrt{7}}{2}\log\frac{\zeta}{\zeta_{\infty}}\right),\label{eq:82}
\end{equation}
where $\mu$ is a constant energy scale. So a simple way to obtain
the asymptotics at large $\zeta$ with massive boundary condition
is to consider a very small variation of $w$,

\begin{equation}
\delta w_{\infty}=\sqrt{\frac{\mu}{\zeta_{\infty}}}2\pi l_{s}^{2}\delta m,
\end{equation}
due to 
\begin{equation}
\lim_{\zeta\rightarrow\infty}w\left(\zeta\right)\rightarrow\frac{1}{\sqrt{\zeta}}.
\end{equation}
Keeping these in hand, then let us investigate the associated variation
in the on-shell action of the flavor brane which is given by recalling
(\ref{eq:73}) (\ref{eq:74})

\begin{align}
\delta S_{\mathrm{D7}} & =\frac{\partial\mathcal{L}}{\partial w^{\prime}}\delta w\bigg|_{\zeta=0}^{\zeta=\zeta_{\infty}}=-T_{7}V_{3}V_{S^{4}}L^{8}\left[\frac{e^{-\phi/4}\zeta^{4}w^{\prime}}{u\left(\rho\right)^{3}\rho^{5}\sqrt{1+w^{\prime2}}}\delta w\right]\bigg|_{\zeta=0}^{\zeta=\zeta_{\infty}}.\label{eq:85}
\end{align}
Notice the relation of the flavor brane energy $E_{f}$ and on-shell
action is $E_{f}=-\frac{S_{\mathrm{D7}}}{V_{3}}$, using (\ref{eq:85}),
the contribution of the massive part to the flavor brane energy (density)
is evaluated as,

\begin{align}
\delta E_{f} & =T_{7}V_{S^{4}}L^{2}\frac{e^{-\phi\left(\zeta_{\infty}\right)/4}\zeta^{2}w^{\prime}\left(\zeta_{\infty}\right)}{\sqrt{1+w^{\prime2}\left(\zeta_{\infty}\right)}}\delta w_{\infty}=\mp c\delta m,\nonumber \\
c & =e^{-\phi\left(\zeta_{\infty}\right)/4}\frac{N_{c}\sqrt{g_{s}N_{c}}}{24\pi^{5/2}}M_{\mu}^{2},
\end{align}
where we have introduced a energy scale $M_{\mu}=2\mu/L^{2}$ and
``$\mp$'' corresponds to the negative/positive mass of the flavor
as it is illustrated in Figure \ref{fig:9}. Afterwards, the total
energy of flavor brane $E_{f}\left(m\right)$ with massive boundary
condition can be obtained by its massless part of energy $E_{f}\left(0\right)$
plus the massive contribution $\delta E_{f}$ as,

\begin{equation}
E_{f}\left(m\right)=E_{f}\left(0\right)+\delta E_{f}.
\end{equation}
Thus the flavor condensate in this system can be obtained as,

\begin{equation}
\left\langle \bar{\psi}\psi\right\rangle =\frac{dE_{f}\left(m\right)}{dm}=\pm c,
\end{equation}
for positive/negative mass.

To close this subsection, let us evaluated the total energy of the
D-brane configuration that $p$ of $N_{f}$ flavor branes extend in
the upper $\left\{ w,\zeta\right\} $ plane while the other $N_{f}-p$
flavor branes extend in the lower $\left\{ w,\zeta\right\} $ plane,
as it is illustrated in Figure \ref{fig:11}. Since the energy of
each flavor brane should be equivalent, for the massless case, the
total energy is given by
\begin{equation}
E_{f}^{tot}\left(0\right)=pE_{f}\left(0\right)+\left(N_{f}-p\right)E_{f}\left(0\right)=N_{f}E_{f}\left(0\right).
\end{equation}
Then for the massive case, suppose the $N_{f}$ flavor branes have
a common mass $m$, the degeneracy between the flavor branes extending
in upper and down lower $\left\{ w,\zeta\right\} $ plane is lifted
for $m\neq0$, therefore we could get the total energy as,

\begin{align}
E_{f}^{tot}\left(m\right) & =p\left[E_{f}\left(0\right)-cm\right]+\left(N_{f}-p\right)\left[E_{f}\left(0\right)+cm\right]\nonumber \\
 & =N_{f}E_{f}\left(0\right)-\left(N_{f}-2p\right)cm.\label{eq:90}
\end{align}
We note that this formula of the total energy in this D-brane configuration
would be useful to study the various phase in the dual theory.

\subsection{The topological phase}

In order to identify the vacuum structure of the D7-branes with the
various phases in the dual theory, we need to give a well-defined
Chern-Simons level in which the Chern-Simons level depends on the
number of the flavor branes. The details to obtain this goal have
been discussed in \cite{key-45,key-46}, so let us briefly outline
the main idea and investigate the phase transition in our holographic
setup. 

\begin{figure}
\begin{centering}
\includegraphics[scale=0.3]{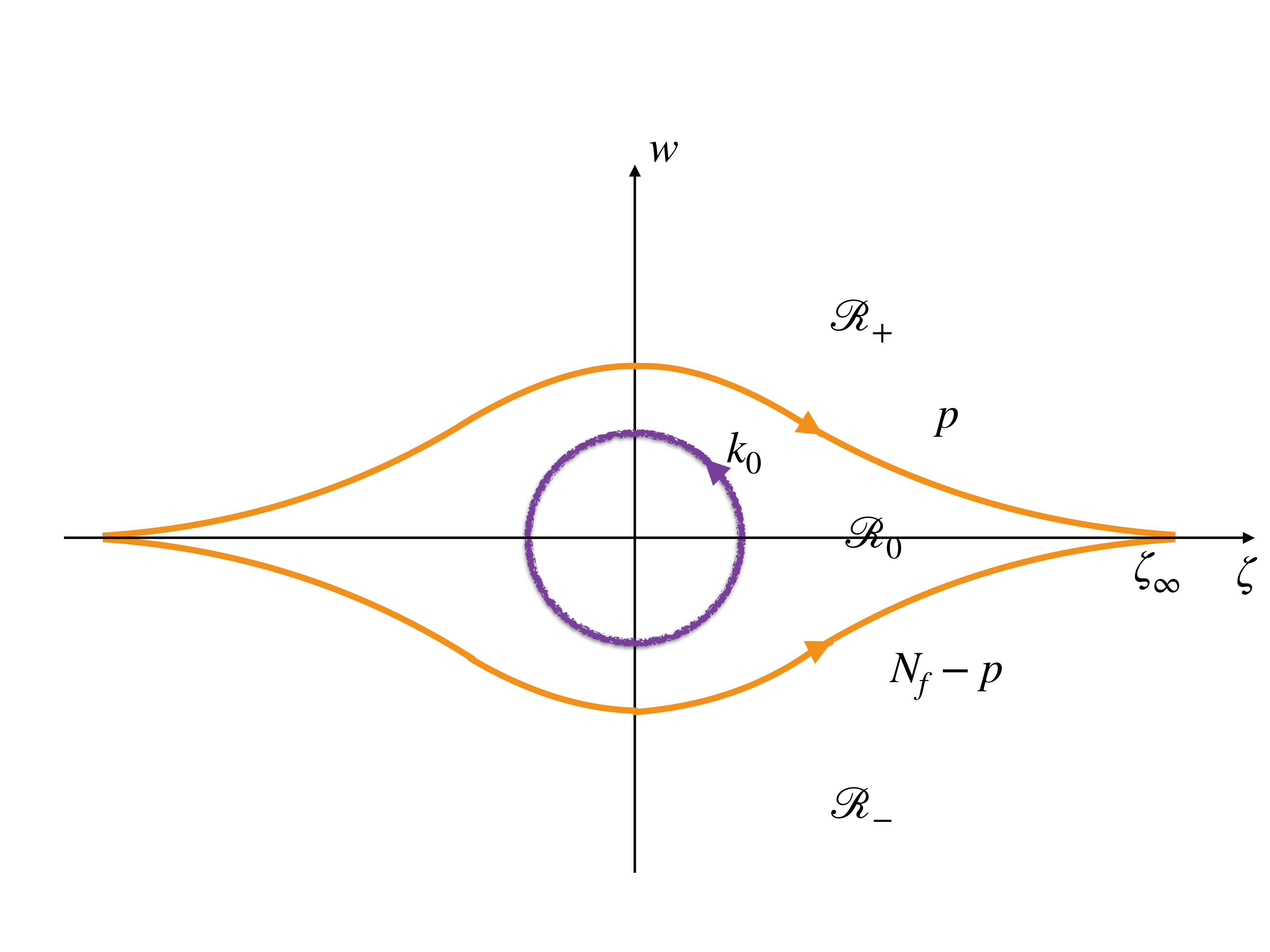}
\par\end{centering}
\caption{\label{fig:11} The D-brane configuration including Chern-Simons (purple)
and flavor (orange) branes in the $\left\{ w,\zeta\right\} $ plane
with massless boundary condition. $p$ of $N_{f}$ flavor branes extend
in the upper plane while the other $N_{f}-p$ flavor branes extend
in the lower plane. The Chern-Simons is located at $u=u_{KK}$ to
minimize its energy as discussed in Section 4.2.}
\end{figure}

First the effective Chern-Simons level $k_{eff}$ according to the
holographic duality is given as,

\begin{equation}
\int_{S^{1}}F_{1}=-k_{eff},
\end{equation}
where $S^{1}$ is a circle whose location is among all other coordinates
except $\left\{ w,\zeta\right\} $. Thus it is a fixed point in the
$\left\{ w,\zeta\right\} $ plane. Then we define the ``$p$ sector'',
that is a D-brane configuration with $p$ flavor branes extending
in the upper $\left\{ w,\zeta\right\} $ plane and $N_{f}-p$ flavor
branes extending in the lower plane. Afterwards, we take into account
the contribution to $k_{eff}$ by counting the number of the orientation
in the $\left\{ w,\zeta\right\} $ plane of the D7-branes. For instance,
the contribution to the effective Chern-Simons number $k_{eff}$ of
D7-branes with counterclockwise/clockwise orientation in the $\left\{ w,\zeta\right\} $
plane is positive/negative respectively, as it is illustrated in Figure
\ref{fig:11} (while we show the massless case, it would be same for
the massive case.). 
 Therefore the effective Chern-Simons level $k_{eff}$ with respected
to the orientation reads,

\begin{equation}
k_{eff}=\begin{cases}
k_{0}-p, & \mathrm{in}\ \mathcal{R}_{+},\\
k_{0}, & \mathrm{in}\ \mathcal{R}_{0},\\
k_{0}+N_{f}-p, & \mathrm{in}\ \mathcal{R}_{-},
\end{cases}
\end{equation}
where $k_{0}$ is the number of the Chern-Simons branes which is given
as $k_{0}=k+p-\frac{N_{f}}{2},k\in\mathbb{Z}$ since both branches
of flavor branes at the intersection point count one-half i.e. $k_{eff}=k_{0}-p+\frac{N_{f}}{2}$.

At low energy, the interpretation of such a D-brane configuration
in holography is that the flavor symmetry $U\left(N_{f}\right)$ spontaneously
breaks down to $U\left(p\right)\times U\left(N_{f}-p\right)$. So
$2p\left(N_{f}-p\right)$ Goldstone bosons are created and the associated
target space is Grassmann,

\begin{equation}
\mathrm{Gr}\left(p,N_{f}\right)=\frac{U\left(N_{f}\right)}{U\left(p\right)\times U\left(N_{f}-p\right)}.
\end{equation}
On the other hand, since the presence of the Chern-Simons leads to
a level/rank duality $U\left(\left|k+p-\frac{N_{f}}{2}\right|\right)_{N}\leftrightarrow SU\left(N\right)_{k+p-N_{f}/2}$,
the dynamics of a $p$ sector at low-energy takes the symmetry,

\begin{equation}
\mathrm{Gr}\left(p,N_{f}\right)\times SU\left(N\right)_{k+p-N_{f}/2},
\end{equation}
in which the vacuum of the dual theory is described by the $N_{f}+1$
sectors via holography. Afterwards, we can investigate the phase diagram
by evaluating the total energy including both flavor and Chern-Simons
branes. Recall (\ref{eq:80}) and (\ref{eq:90}), the total energy
including flavor and Chern-Simons branes is collected as,

\begin{align}
E_{\mathrm{vac}}\left(a\right) & =E_{f}^{tot}\left(m\right)+k_{0}E_{\mathrm{CS}}\nonumber \\
 & =N_{f}E_{f}\left(0\right)-\left(N_{f}-2p\right)cm+\left(k+p-\frac{N_{f}}{2}\right)E_{\mathrm{CS}}.\label{eq:95}
\end{align}
The phase diagram can be obtained by minimizing the energy $E_{\mathrm{vac}}\left(a\right)$
in (\ref{eq:95}). To find the dependence on the anisotropy (denoted
by $a$), we require that the value of $a$ is fixed when we minimize
$E_{\mathrm{vac}}\left(a\right)$. Then the results are collected
as, for $k>N_{f}/2$,

\begin{equation}
E_{\mathrm{vac}}=\begin{cases}
\left(k-N_{f}/2\right)E_{\mathrm{CS}}, & m<m^{*},\ \ \ \ SU\left(N_{c}\right)_{k-N_{f}/2},\\
\left(k+N_{f}/2\right)E_{\mathrm{CS}}-2N_{f}cm, & m>m^{*},\ \ \ \ SU\left(N_{c}\right)_{k+N_{f}/2},
\end{cases}
\end{equation}
where $SU\left(N_{c}\right)_{k\pm N_{f}/2}$ refers to the symmetry
group of the corresponding topological phase. And for $k<N_{f}/2$,
the associated topological phase and free energy are collected as,

\begin{equation}
E_{\mathrm{vac}}=\begin{cases}
\left(N_{f}/2-k\right)E_{\mathrm{CS}}, & \ \ \ m<-m^{*},\ \ \ \ \ \ \ SU\left(N_{c}\right)_{k-N_{f}/2},\\
2\left(k-N_{f}/2\right)cm, & -m^{*}<m<m^{*},\ \ \ \ \ \ \ \mathrm{Gr}\left(p,N_{f}\right),\\
\left(N_{f}/2+k\right)E_{\mathrm{CS}}-2N_{f}cm, & \ \ \ \ m>m^{*},\ \ \ \ \ \ \ \ SU\left(N_{c}\right)_{k+N_{f}/2},
\end{cases}
\end{equation}
where $m^{*}$ refers to the critical value of the mass when the phase
transition occurs, given as,

\begin{equation}
m^{*}=\frac{E_{\mathrm{CS}}}{2c}=\frac{3}{16}\frac{\lambda^{1/2}\pi^{1/2}M_{KK}^{3}}{M_{\mu}^{2}}\left(1+\frac{2-\log16}{4}\frac{a^{2}}{M_{KK}^{2}}\right)+\mathcal{O}\left(a^{4}\right).
\end{equation}
Here a noteworthy feature is that the critical mass may become vanished
if the axion field or the anisotropy in the bulk becomes very non-negligible
i.e $a$ becomes sufficiently large. In this sense the Grassmann phase
$\mathrm{Gr}\left(p,N_{f}\right)$ would not exist which seemingly
means the broken flavor symmetry is restored. Although our setup may
be exactly valid only for small anisotropy, this result is instructively
suggestive to study the anisotropic behavior of the metastable vacua
in QCD\textsubscript{3} via holography.

\section{Summary and discussion}

In this work, we construct the anisotropic black D3-brane solution
in IIB supergravity \cite{key-9} then obtain the anisotropic bubble
configuration for QCD\textsubscript{3} with a Chern-Simons term due
to the presence of the axion field. The analytical formulas for the
the background geometry is available since the dual theory is exactly
three-dimensional theory in the compactification limit, as it is expected.
With this analytical bulk geometry, we investigate the ground-state
energy density, quark potential, entanglement entropy and baryon vertex
in the dual theory according to the AdS/CFT dictionary. Technically,
we consider small anisotropy to avoid the difficulty in our numerical
calculation. Then all the results show the dependence of the axion
field or the anisotropy in bulk as it is expected. Afterwards we introduce
various probe D7-branes as flavor and Chern-Simons branes to include
flavor matters and topological numbers in the dual theory. By examining
the embedding functions and counting the orientation of these D7-branes,
we obtain the vacuum energy associated to the corresponding effective
Chern-Simons level, hence the phase transition can be achieved by
comparing the various vacuum energies. 

To close this work, let us give some comments about this project.
Due to the presence of the axion in bulk, the quark potential and
entanglement entropy are shifted as some holographic studies in four-dimensional
QCD with an axion e.g. \cite{key-15,key-23}. However our work additionally
implies the quark tension and the potential phase transition illustrated
in the behavior of the entanglement entropy could be destroyed in
the presence of strong anisotropy. These can be found in Section 3.2
and Section 3.3: when the anisotropy increases, we can see the quark
tension trends to become vanished and there would not be a critical
value of $l^{\perp,||}$ satisfying the entanglement entropy $\Delta S^{\perp,||}=0$.
As the entanglement entropy could be a tool to characterize the confinement
\cite{key-36,key-37,key-38,key-39}, this behavior implies there would
be no phase transition for $a\gg1$ i.e. no confinement for strong
anisotropy. Besides, the baryon vertex also reveals the unwrapped
trend when the anisotropy becomes large and the ``unwrapped baryon
vertex'' also means deconfinement \cite{key-44}. In a word, this
holographic approach shows us the confinement can not maintain in
an extremely anisotropic situation. Interestingly, this conclusion
is in agreement with the fact that the QGP is anisotropic and deconfined,
so it may provide a holographic way to understand the features of
the strongly coupled matter with anisotropy.

On the other hand, the dependence on the axion or the anisotropy of
the critical mass in the topological phase transition would also be
a parallel computation to the D3-(D-1) approach \cite{key-45} and
the extension of \cite{key-46} by including an dynamical axion field.
Moreover, as the critical mass trends to be vanished when the anisotropy
increases, it means the Grassmann target space would not exist. Namely
the broken flavor symmetry would be restored if the anisotropy becomes
sufficiently large. We note that this behavior is also illustrated
in Figure \ref{fig:8}. As we can see in Figure \ref{fig:8} the flavor
branes in the upper and lower branches trend to be coincident if $a$
increases greatly, i.e. the flavor symmetry $U\left(p\right)\times U\left(N_{f}-p\right)$
would be restored to $U\left(N_{f}\right)$ if $a\rightarrow\infty$.
Accordingly, this behavior implies the flavor symmetry, which is related
to the chiral symmetry, would be restored when the anisotropy is extremely
strong. 

Combine the above together, this holographic system reveals a potential
conclusion that is the confinement will not maintain and the flavor
symmetry (or probably chiral symmetry) would be restored in an extremely
anisotropic situation. This conclusion adheres to intuition, because
if the anisotropy becomes very large for a fixed $M_{KK}$ as $a\gg M_{KK}$,
the dual theory depending on $a$ would include modes above the scale
$M_{KK}$ thus the dual theory is decompactified and non-confining
according to \cite{key-35,key-36}. Remarkably, all the analyses are
exactly coincident with the characteristic properties of QGP, particularly
it is usually to be treated as the fundamental assumption to study
the deconfined matter in holography \cite{key-57,key-58,key-59,key-60}.
Therefore, while we can only work out numerically the case that the
anisotropy is small in this project, our framework would be very instructive
to study QCD and Chern-Simons theory with anisotropy.

\section*{Acknowledgements}

We would like to thank Niko Jokela for helpful discussion. This work
is supported by the National Natural Science Foundation of China (NSFC)
under Grant No. 12005033, the research startup foundation of Dalian
Maritime University in 2019 under Grant No. 02502608 and the Fundamental
Research Funds for the Central Universities under Grant No. 3132022198.

\section*{Appendix: The analytical formulas for the anisotropic background}

In the high temperature limit $T\rightarrow\infty$, $\delta t_{E}\rightarrow0$.
The functions $\mathcal{F},\mathcal{B},\phi$ presented in the anisotropic
black brane background (\ref{eq:2}) can be written as a series of
$a$ up to $\mathcal{O}\left(a^{2}\right)$ as \cite{key-9},

\begin{align}
\mathcal{F}\left(u\right) & =1-\frac{u^{4}}{u_{H}^{4}}+a^{2}\hat{\mathcal{F}}_{2}\left(u\right)+\mathcal{O}\left(a^{4}\right),\nonumber \\
\mathcal{B}\left(u\right) & =1+a^{2}\hat{\mathcal{B}}_{2}\left(u\right)+\mathcal{O}\left(a^{4}\right),\nonumber \\
\phi\left(u\right) & =a^{2}\hat{\phi}_{2}\left(u\right)+\mathcal{O}\left(a^{4}\right),\tag{A-1}\label{eq:99}
\end{align}
where

\begin{align}
\hat{\mathcal{F}}_{2}\left(u\right) & =\frac{1}{24u_{H}^{2}}\left[8u^{2}\left(u_{H}^{2}-u^{2}\right)-10u^{4}\log2+\left(3u_{H}^{4}+7u^{4}\right)\log\left(1+\frac{u^{2}}{u_{H}^{2}}\right)\right],\nonumber \\
\hat{\mathcal{B}}_{2}\left(u\right) & =-\frac{u_{H}^{2}}{24}\left[\frac{10u^{2}}{u_{H}^{2}+u^{2}}+\log\left(1+\frac{u^{2}}{u_{H}^{2}}\right)\right],\nonumber \\
\hat{\phi}_{2}\left(u\right) & =-\frac{u_{H}^{2}}{4}\log\left(1+\frac{u^{2}}{u_{H}^{2}}\right).\tag{A-2}\label{eq:100}
\end{align}
We note that (\ref{eq:99}) and (\ref{eq:100}) is in fact a series
of $u_{H}a$ (or equivalently $a/T$), so the high temperature limit
exactly refers to the case $T\gg a$ in \cite{key-9} which corresponds
to $u_{H}a\ll1$ or $a/T\ll1$ in (\ref{eq:99}) (\ref{eq:100}).
In the black brane background (\ref{eq:2}), $u_{H}$ refers to the
horizon. In the bubble background (\ref{eq:6}), the double wick rotation
reduces to the replacement $T\rightarrow M_{KK}/\left(2\pi\right),\delta t_{E}\rightarrow\delta z$
in the black brane solution. Note that $u_{H}$ is replaced by $u_{KK}$
in the bubble background (\ref{eq:6}) since $u_{KK}$ refers to the
bottom of the bulk instead of a horizon as it is illustrated in Figure
\ref{fig:1}. And the formulas of $\mathcal{F},\mathcal{B},\phi$
remain while we replace $u_{H}$ by $u_{KK}$. Clearly the high temperature
limit in the black brane solution corresponds to the limit of dimension
reduction in the bubble solution i.e. the limit for that the compactified
direction $z$ shrinks to zero in (\ref{eq:6}) i.e. $\delta z\rightarrow0$,
or equivalently $M_{KK}\rightarrow\infty$. 

On the other hand, the dual theory on the $N_{c}$ D3-branes in the
bubble background (\ref{eq:6}) is effectively three-dimensional below
the energy scale $M_{KK}$. So if we take $M_{KK}\rightarrow\infty$,
the dual theory would become exactly three-dimensional theory. Therefore,
the analytical formulas (\ref{eq:99}) (\ref{eq:100}) for functions
$\mathcal{F},\mathcal{B},\phi$ can always be employed by replacing
$u_{H}\rightarrow u_{KK}$ in the bubble background (\ref{eq:6})
since the dual theory is always expected to be exactly three-dimensional
theory, which means $M_{KK}\rightarrow\infty$ is always expected
even if $a$ becomes large but fixed. In a word, using (\ref{eq:99})
(\ref{eq:100}) in the bubble background means the dual theory is
exactly three-dimensional theory. In this sense, we believe our analysis
in this work with (\ref{eq:99}) (\ref{eq:100}) is also valid for
large $a$ under $M_{KK}\rightarrow\infty$.

\end{document}